\documentclass[aps,showpacs,11pt]{revtex4}
\usepackage{dcolumn}
\usepackage{graphicx}
\usepackage{amsmath}
\usepackage{amsfonts}
\usepackage{amssymb}
\usepackage{psfrag}
\usepackage{wrapfig}
\usepackage{subfigure}
\usepackage{makeidx}
\usepackage{bm}
\usepackage{epsf}
\usepackage{cases}
\usepackage{hyperref}
\usepackage{multirow}
\makeatletter

\newcommand{\Rmnum}[1]{\expandafter\@slowromancap\romannumeral #1@}
\makeatother

\begin{document}

\title{Behavior of quasinormal modes and Van der Waals-like phase
transition of charged AdS black holes in massive gravity}

\author{De-Cheng Zou$^{1}$\footnote{Email:dczou@yzu.edu.cn},
Yunqi Liu$^{2}$\footnote{Email:liuyunqi@hust.edu.cn}
and Ruihong Yue$^{1}$ \footnote{Email:rhyue@yzu.edu.cn}}

\affiliation{ $^{1}$Center for Gravitation and Cosmology,
College of Physical Science and Technology, Yangzhou University, Yangzhou 225009, China\\
$^{2}$School of Physics, Huazhong University of Science and Technology, Wuhan 430074, China}

\date{\today}

\begin{abstract}
\indent

In this work, we utilize the quasinormal modes (QNMs)
of a massless scalar perturbation to probe the Van der Waals-like
small and large black holes (SBH/LBH) phase transition of
charged topological Anti-de Sitter (AdS) black holes in four-dimensional
massive gravity. We find that the signature of this SBH/LBH phase transition
is detected in the isobaric as well as in the isothermal process.
This further supports the idea that the QNMs can be an efficient tool to
investigate the thermodynamical phase transition.

\end{abstract}


\maketitle

\section{Introduction}
\label{intro}

Einstein's general relativity introduces gravitons as massless spin-2
particles \cite{Gupta:1954zz,Weinberg:1965rz,Feynman:1996kb}.
However, understanding the quantum behavior of gravity could be related to the possible
mass of the graviton. This Einstein theory, modified at large distances in massive
gravity provides a possible explanation for the accelerated expansion of the Universe
that does not require any dark energy. Actually, the massive gravity and its extensions, such as
bimetric gravity, can yield cosmological solutions which do display late-time
acceleration in agreement with observations \cite{Hassan:2011zd,DAmico:2011eto,Akrami:2012vf,Akrami:2015qga}.
Very recently, the LIGO collaboration reporting the discovery of gravitational wave asserted
that \cite{Abbott:2016blz} ``assuming a modified dispersion relation for gravitational
waves, our observations constrain the Compton wavelength of the graviton to
be $\lambda_g>10^{13} km$, which could be interpreted
as a bound on the graviton mass $m_g<1.2\times10^{-22}eV/c^2$''.
In order to have massive graviton,
the first attempt for constructing a massive theory
was in the work of Fierz and Pauli \cite{Fierz:1939ix} which was done in
the context of linear theory. Unfortunately this theory possesses so-called van Dam,
Veltman and Zakharov discontinuity problem. The resolution to this problem was
Vainshtein's mechanism, which requires the
system to be considered in nonlinear framework.
As is now well known, it usually brings about Boulware-Deser ghost \cite{Boulware:1973my}
by adding generic mass terms for the graviton on the nonlinear level.
Subsequently, a nonlinear massive gravity theory was proposed by de Rham, Gabadadze
and Tolley (dRGT) \cite{deRham:2010ik,deRham:2010kj},
where the mass terms are added in a specific way to ensure that
the corresponding equations of motion are at most second order differential equations so
that the Boulware-Deser ghost is eliminated.
Later, spherically symmetric black hole solutions were
constructed in the dRGT massive gravity \cite{Vegh:2013sk,Ghosh:2015cva,Do:2016abo,Li:2016fbf},
including its extension in terms of electric
charge  \cite{Cai:2012db,Berezhiani:2011mt,Hendi:2015hoa},
black string \cite{Tannukij:2017jtn}, BTZ like black holes \cite{Hendi:2016hbe,Hendi:2016pvx}
and some other solutions with higher curvature correction terms
\cite{Hendi:2015pda,Meng:2016its}. This program goes beyond solution constructions
in the dRGT massive gravity and focuses on the investigation holographic implications \cite{Zeng:2014uoa,Hendi:2016uni,Baggioli:2014roa,Ge:2014aza,Hu:2015dnl,Sadeghi:2015vaa},
discussing the thermodynamical properties \cite{Cai:2014znn,Hendi:2015bna,Hendi:2016yof,Adams:2014vza}
and calculating QNMs under massless scalar perturbations for the BTZ like
black hole \cite{Prasia:2016esx}.

The thermodynamical phase transition of a black hole is always a hot topic
in black hole physics. It may shed light on
the understanding of the relation between gravity and thermodynamics.
Recently, thermodynamics of AdS black holes has been generalized
to the extended phase space where the cosmological constant is treated
as the pressure of the black hole \cite{Caldarelli:1999xj,Kastor:2009wy,Lu:2012xu}.
A particular emphasis has been put on the study of the black hole phase
transitions in AdS spacetime in Ref.\cite{Kubiznak:2012wp}, which asserted the
analogy between the Van der Waals liquid-gas system
behavior and the charged AdS black hole.
Subsequently a broad range of thermodynamic behaviors have been discovered,
including reentrant phase transitions and more general Van der Waals behavior  \cite{Gunasekaran:2012dq,Hendi:2015hgg,Hendi:2012um,Zhao:2013oza,
Zou:2013owa,Cai:2013qga,Mo:2014qsa,Zou:2014mha,Xu:2014tja,Xu:2014kwa,
Rajagopal:2014ewa,Hennigar:2015esa,Hendi:2016usw,Zhang:2014jfa,
Dehghani:2014caa,Hansen:2016ayo,Sadeghi:2016dvc,Wei:2015iwa,Wei:2015ana,
Liang:2017rng,Frassino:2014pha,Wei:2014hba,Lan:2017yia,Cheng:2016bpx,Altamirano:2014tva,
Zeng:2016fsb,Xu:2013zea,Mo:2016ndm,Belhaj:2014eha,Hendi:2015cqz,B.:2015koa}.
Recently, some investigations od the thermodynamics of AdS
black holes in the massive gravity showed generalization to the extended phase
space \cite{Xu:2015rfa,Hendi:2017fxp,Zeng:2015tfj,Prasia:2016fcc,
Zou:2016sab,Hendi:2015hoa,Ning:2016usb},
including the higher curvature terms \cite{Hendi:2015pda,Hendi:2016yof,Hendi:2015bna}.

For a long time, thermodynamical phase transitions of the black hole are supposed
to detect by some observational signatures. Considering
that QNMs of dynamical perturbations are characteristic
issues of black holes \cite{Nollert:1999ji,Kokkotas:1999bd,Konoplya:2011qq},
it is expected that black hole phase transitions can be reflected in the
dynamical perturbations in the surrounding geometries of black holes through
frequencies and damping times of the oscillations. Moreover,
the QNM frequencies of AdS black holes have direct interpretation
in terms of the dual conformal field theory CFT \cite{Cardoso:2013pza,Cardoso:2003cj,Warnick:2013hba,Konoplya:2002ky,Konoplya:2008rq,Li:2016kws}.
A lot of discussions have been focused on this topic and more and
more evidence has been found between thermodynamical phase transitions
and dynamical perturbations. See for examples \cite{Rao:2007zzb,He:2010zb,Miranda:2008vb,
He:2008im,Pan:2011hj,Berti:2008xu,Shen:2007xk,Koutsoumbas:2006xj,Zou:2014sja,
Zangeneh:2017rhc,Cai:2011qm}.
In the extended phase space, we have recovered the deep relation between the
dynamical perturbation and the Van der Waals-like SBH/LBH phase transition
in the four-dimensional Reissner-Nordstr$\ddot{o}$m-Anti de Sitter (RN-AdS) black holes
with spherical horizon $(k=1)$ \cite{Liu:2014gvf}. Later, matters have been
generalized to higher-dimensional RN-AdS black holes \cite{Chabab:2016cem} ,
including time-domain profiles \cite{Chabab:2017knz}, and higher-dimensional
charged black holes in the presence of Weyl coupling \cite{Mahapatra:2016dae}.

It is necessary to point out that in four-dimensional dRGT massive gravity,
there always exists so-called Van der Waals-like SBH/LBH phase transition
for the charged AdS black holes when the horizon topology is spherical $(k=1)$,
Ricci flat $(k=0)$ or hyperbolic $(k=-1)$ \cite{Xu:2015rfa}.
In particular, this phenomenon rarely occurs, since this Van der Waals-like
 SBH/LBH phase transition was usually recovered in a
variety of spherical horizon black hole backgrounds.
Motivated by these results,  in this paper we find crucial and well justified to
reconsider the charged topological AdS black hole in
four-dimensional dRGT massive gravity.
We further use the QNM frequencies of a massless scalar perturbation to
probe the Van der Waals-like SBH/LBH phase transitions of charged topological
black holes $(k=0, k=\pm1)$, respectively.

This paper is organized as follows. In Sect.~\ref{2s}, we will review
the Van der Waals-like SBH/LBH phase transition of charged topological
AdS black holes in four-dimensional massive gravity. In Sect.~\ref{3s},
we will disclose numerically that the phase transition can be reflected
by the QNM frequencies of dynamical perturbations. We end the paper with
conclusions and discussions in Sect.~\ref{4s}.

\section{Phase transition of charged topological AdS black hole in massive gravity}
\label{2s}
We start with the action of four-dimensional massive gravity in the
presence of a negative cosmological constant \cite{Cai:2014znn}
\begin{eqnarray}
{\cal I}=\frac{1}{16\pi}\int{d^4x\sqrt{-g}}\left[R-2\Lambda
-\frac{1}{4}F_{\mu\nu}F^{\mu\nu}+m^2\sum_i^{4}c_{i}{\cal U}_{i}(g,f)\right],\label{action}
\end{eqnarray}
where $f$ is a fixed symmetric tensor usually called the reference metric,
$c_i$ are constants, $m$ is the mass parameter related to the graviton mass,
and $F_{\mu\nu}$ is the Maxwell field strength defined as
$F_{\mu\nu}=\partial_{\mu}A_{\nu}-\partial_{\nu}A_{\mu}$ with vector potential $A_{\mu}$.
Moreover, ${\cal U}_{i}$ are symmetric polynomials of the eigenvalues
of the $4\times 4$ matrix ${\cal K}^\mu_{\nu}\equiv\sqrt{g^{\mu\alpha}f_{\alpha\nu}}$
\begin{eqnarray}
{\cal U}_{1}&=&[\cal K],\nonumber\\
{\cal U}_{2}&=&[{\cal K}]^2-[{\cal K}^2],\nonumber\\
{\cal U}_{3}&=&[{\cal K}]^3-3[{\cal K}][{\cal K}^2]+2[{\cal K}^3],\nonumber\\
{\cal U}_{4}&=&[{\cal K}]^4-6[{\cal K}^2][{\cal K}]^2
+8[{\cal K}^3][{\cal K}]+3[{\cal K}^2]^2-6[{\cal K}^4].\nonumber\\\label{ac}
\end{eqnarray}
The square root in ${\cal K}$ is understood as the matrix square root, ie.,
$(\sqrt{A})^{\mu}_{~\nu}(\sqrt{A})^{~\nu}_{\lambda}=A^\mu_{~\lambda}$, and the
rectangular brackets denote traces $[{\cal K}]={\cal K}^{\mu}_{~\mu}$.

The action admits a static black hole solution with metric
\begin{eqnarray}
ds^2=-f(r)dt^2+\frac{1}{f(r)}dr^2+r^2h_{ij}dx^idx^j,\label{metric}
\end{eqnarray}
where the coordinates are labeled $x^{\mu}=(t, r, x^1, x^2)$ and $h_{ij}$ describes
the two-dimensional hypersurface with constant scalar curvature $2k$.
The constant $k$ characterizes the geometric property of a
hypersurface, which takes values $k=0$ for the flat case, $k=-1$ for
negative curvature and $k=1$ for positive curvature, respectively.
In a four-dimensional situation, we have ${\cal U}_{3}={\cal U}_{4}=0$.
Then the solution of charged topological AdS
black hole is given by \cite{Cai:2014znn}
\begin{eqnarray}
f(r)=k+\frac{8\pi P r^2}{3}-\frac{m_0}{r}+\frac{q^2}{4r^2}
+\frac{c_0c_1m^2r}{2}+c_0^2c_2m^2,\label{solution}
\end{eqnarray}
where $P$ equals to $-\frac{\Lambda}{8\pi}$.
Moreover, the parameters $m_0$ and $q$
are related to mass and charge of black hole
\begin{eqnarray}
M=\frac{V_2}{8\pi}m_0,\quad Q=\frac{V_2}{16\pi}q.\nonumber
\end{eqnarray}
Here $V_2$ is the volume of space spanned by coordinates $x^i$.
When $m\rightarrow0$, the solution (\ref{solution})
reduces to the RN-AdS black hole.

The reference metric now can have a special choice
\begin{eqnarray}
f_{\mu\nu}=diag(0,0,c_0^2h_{ij}).\label{remetric}
\end{eqnarray}
Without loss of generality
we have set $c_0=1$ in our following discussions.

In terms of the radius of the horizon $r_+$, the mass $M$,
Hawking temperature $T$, entropy $S$ and electromagnetic potential $\Phi$
of the black holes can be written as
\begin{eqnarray}
M&=&\frac{V_2 r_+}{8\pi}\left(k+c_{2}m^2+\frac{q^2}{4r_+^2}+\frac{8\pi P r_+^2}{3}
+\frac{c_{1}m^2 r_{+}}{2}\right),\nonumber\\
T&=&-\frac{q^2}{16\pi r_+^3}+\frac{k+c_2m^2}{4\pi r_+}+2 Pr_{+}
+\frac{c_{1}m^2}{4\pi},\nonumber\\
S&=&\frac{V_2}{4} r_+^2,\qquad \Phi=\frac{q}{r_+}\label{eq:5a}
\end{eqnarray}
In the extended phase space, the black hole mass $M$ is considered as the
enthalpy rather than the internal energy of the
gravitational system \cite{Kastor:2009wy}.

From Eq.~(\ref{eq:5a}), the equation of state of the black hole can be obtained
\begin{eqnarray}
P=\frac{T}{2r_+}-\frac{c_{1}m^2}{8\pi r_+}-\frac{k+c_{2}m^2}{8\pi r_+^2}
+\frac{q^2}{32\pi r_+^4}.\label{eos}
\end{eqnarray}
To compare with Van der Waals fluid equation in four dimensions, we can translate
the ``geometric'' equation of state to a physical one by identifying the specific
volume $v$ of the fluid with horizon radius of black hole as $v=2r_+$.

As usual, a critical point occurs when $P$ has an inflection point
\begin{eqnarray}
\frac{\partial P}{\partial r_+}\Big|_{T=T_c, r_+=r_c}
=\frac{\partial^2 P}{\partial r_+^2}\Big|_{T=T_c, r_+=r_c}=0,\label{eq:15a}
\end{eqnarray}
which leads to
\begin{eqnarray}
r_c=\frac{\sqrt{6}q}{2\sqrt{k+c_{2}m^2}}, \quad
T_c=\frac{2\left(k+c_{2}m^2\right)^{3/2}}{3\sqrt{6}\pi q}
+\frac{c_{1}m^2}{4\pi},\quad
P_c=\frac{(k+c_{2}m^2)^2}{24\pi q^2}.
\end{eqnarray}
Evidently the critical behavior occurs when
$k+c_{2}m^2>0$, which is a joint effect of horizon topology $k$ and $c_2m^2$.
Previous thermodynamical discussions for RN-AdS black holes
show that the Van der Waals-like SBH/LBH only occurs for spherical horizon
topology $k=1$. The graviton mass significantly modifies this behavior and
a non-zero $m$ admits that possibility of critical behavior for $k\neq1$.
In addition, it has been shown \cite{Cai:2014znn} that when $k+c_{2}m^2>0$,
the small and large black hole phases are both
locally thermodynamically stable because the corresponding heat capacities
are always positive\footnote{We thank Hai-Qing Zhang for pointing this out.}.

The equilibrium thermodynamics is governed by the
Gibbs free energy, $G=G(T, P, q)$, which obeys the thermodynamic
relation $G=M-TS$. For later discussions, it is convenient to rescale the Gibbs free
energy in the following way: $g=\frac{4\pi}{V_2}G$. Then $g$ reads
\begin{eqnarray}
g=\frac{3q^2}{16r_+}+\frac{\left(k+c_{2}m^2\right)r_+}{4}
-\frac{2\pi P r_+^3}{3}.\label{freeE}
\end{eqnarray}
Here $r_+$ is understood as a function of pressure and temperature, $r_+=r_+(P,T)$, via
the equation of state (\ref{eos}).

\section{Perturbations of charged topological AdS black hole in massive gravity}
\label{3s}

Now we study the evolution of a massless scalar field perturbation in the surrounding
geometry of these charged topological AdS black holes.

A massless scalar field $\Psi(r,t,\Omega)=\phi(r)e^{-i\omega t}Y_{lm}(\Omega)$,
obeys the Klein-Gordon equation
\begin{eqnarray}
\nabla_{\mu}^{2}\Psi(r,t,\Omega)=\frac{1}{\sqrt{-g}}\partial_\mu\left(\sqrt{
-g}g^{\mu\nu}\partial_\nu\Psi(r,t,\Omega)\right)=0,\label{KG}
\end{eqnarray}
where $Y_{lm}(\Omega)$ is a normalizable harmonic function on the 2-dimensional hypersurface.
In particular, the Laplace operator on $\Omega$ yields
\begin{eqnarray}
\nabla^{2}_{\Omega}Y_{lm}(\Omega)=-\kappa^2Y_{lm}(\Omega),\label{KG1}
\end{eqnarray}

It is necessary to point out that the eigenvalue $\kappa^2$ usually gets
different values in consideration of different horizon topologies.
For the spherical $(k=1)$ and flat $(k=0)$ topology,
the eigenvalue $\kappa^2$ can be zero. Then
the radial function $\phi(r)$ obeys
\begin{eqnarray}
\phi''(r)+\left(\frac{f'(r)}{f(r)}+\frac{2}{r}\right)\phi'(r)
+\frac{\omega^2\phi(r)}{f(r)^2}=0,\label{KG12}
\end{eqnarray}
where $\omega$ are complex numbers $\omega=\omega_r + i\omega_{im}$, corresponding to
the QNM frequencies of the oscillations describing the perturbation.
For the hyperbolic horizon topology$(k=-1)$, the
eigenvalue $\kappa^2$ of the Laplace operator on $\Omega$ cannot be zero \cite{Alsup:2008fr,Gonzalez:2012xc,Balazs:1986uj,Becar:2012bj}, and
is given by $\frac{1}{4}+\xi^2$, where $\xi=L_{\Omega}(L_{\Omega}+1)$,
$L_{\Omega}=0,1,2,...$\cite{Becar:2015kpa}.
Then the radial function $\phi(r)$ obeys the following differential
equation:
\begin{eqnarray}
\phi''(r)+\left(\frac{f'(r)}{f(r)}+\frac{2}{r}\right)\phi'(r)
+\left(\frac{\omega^2}{f(r)}-\frac{\kappa^2}{r^2}\right)\frac{\phi(r)}{f(r)}=0.\label{KG21}
\end{eqnarray}

Here we define $\phi(r)$ as $\varphi(r)exp[-i\int\frac{\omega}{f(r)}dr]$,
where the $exp[-i\int\frac{\omega}{f(r)}dr]$ asymptotically approaches to ingoing wave
near horizon, then Eqs.~(\ref{KG12})(\ref{KG21}) become
\begin{eqnarray}
\varphi''(r)+\varphi'(r)\left(\frac{f'(r)}{f(r)}-\frac{2i\omega}{f(r)}+\frac{2}{r}\right)
-\frac{2i\omega}{r f(r)}\varphi(r)=0,\qquad \qquad \qquad k=0,1 \label{omegaKG1}
\end{eqnarray}
and
\begin{eqnarray}
\varphi''(r)+\varphi'(r)\left(\frac{f'(r)}{f(r)}-\frac{2i\omega}{f(r)}+\frac{2}{r}\right)
-\left(2i\omega+\frac{\kappa^2}{r}\right)\frac{\varphi(r)}{r f(r)}=0,\qquad k=-1.\label{omegaKG2}
\end{eqnarray}
In this paper, we only consider $\xi=0$, namely $\kappa^2=1/4$ for $k=-1$.

We are going to study whether the signature of Van der Waals-like SBH/LBH
phase transition of charged topological AdS black holes can be
reflected by the dynamical QNMs behavior in the massless scalar perturbation.
For Eqs.~(\ref{omegaKG1}) and (\ref{omegaKG2}), we have $\varphi(r)=1$ in the limit of $r\rightarrow r_+$.
At the AdS boundary $(r\rightarrow\infty)$, we need $\varphi(r)=0$.
Under these boundary conditions, we will numerically solve Eqs.~(\ref{omegaKG1}) and (\ref{omegaKG2})
separately to find QNM frequencies by adopting the shooting method.
In the context of the Van der Waals phase transition picture, the dynamical
perturbations in the isobaric process and isothermal process will be discussed.
In our following numerical computations we will set $q=2$, $m=1$, $c_1=0.05$
and $c_2=2$.

\subsection{Isobaric phase transition}

\begin{figure}
\centering
\includegraphics[width=2.7in]{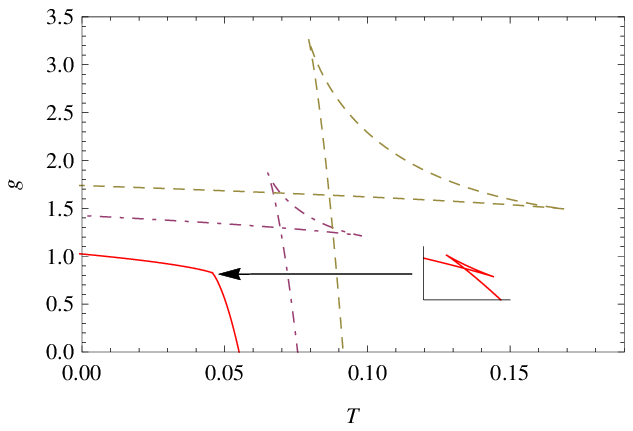}
\hfill%
\includegraphics[width=2.8in]{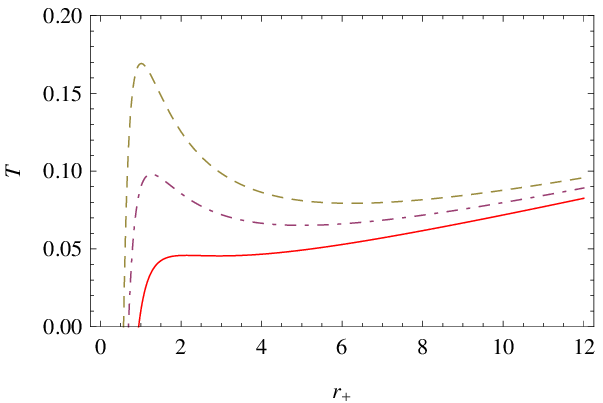}
\caption{The $g-T$(left panel) and $T-r_+$ (right panel) diagrams for $P<P_c$.
The three lines correspond to $k=1$(dashed line), $k=0$(dotdashed line)
and $k=-1$(solid line).}\label{fig1}
\end{figure}

Due to the pressure $P$ (or $l$) being fixed in this case, the black hole horizon $r_+$
is the only variable in the system. The behavior of an isobar with different horizon topologies
are plotted in Fig.~\ref{fig1}. For $P<P_c$, the oscillating part displays
the occurrence of an SBH/LBH phase transition in the system and the Gibbs free energy depicts
a swallow tail behavior, also signaling a first-order SBH/LBH phase transition.
Here the intersection point indicates the coexistence of two phases in equilibrium.
The critical pressure $P_c$ is
obtained by $\frac{\partial T}{\partial r_+}=\frac{\partial^2 T}{\partial r_+^2}=0$.

%%%%%%%%%%%%%%%%%%%%%%%%%%%%%%%%%%%%%%%%%%%%%%%%%%%%%%%%%%%%%%%%%%%%%%%%%%%

\begin{figure}
\centering
\includegraphics[width=2in]{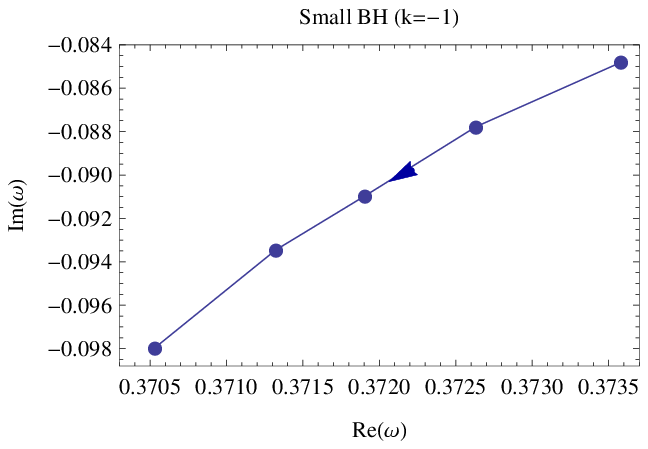}
\hfill%
\includegraphics[width=2in]{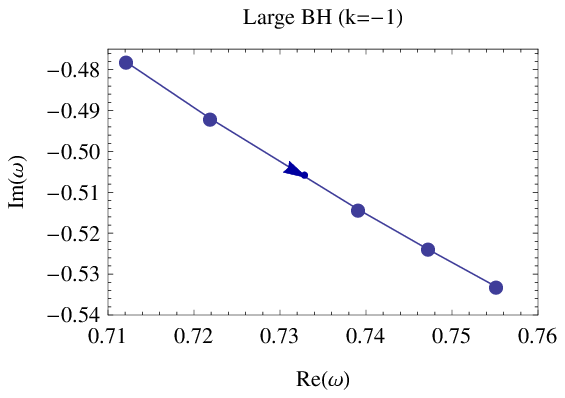}
\hfill%
\includegraphics[width=2in]{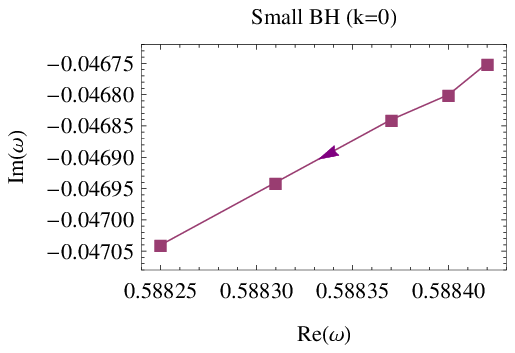}
\hfill%
\includegraphics[width=2in]{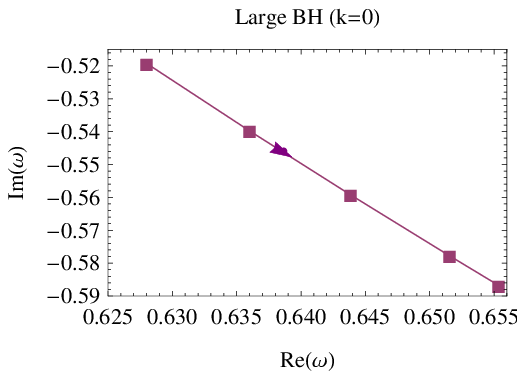}
\hfill%
\includegraphics[width=2in]{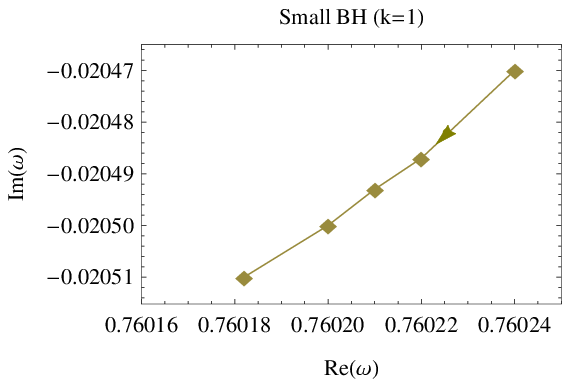}
\hfill%
\includegraphics[width=2in]{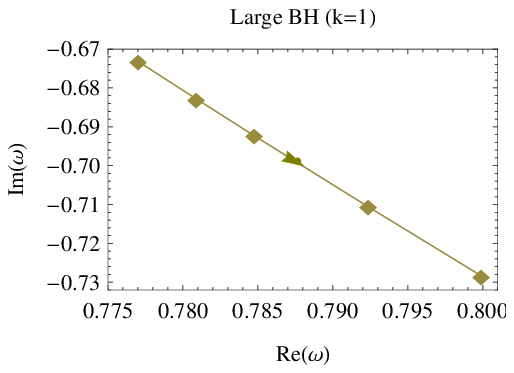}
\caption{The behaviors of QNM frequencies for large and small black holes
in the isobaric process. The arrow indicates the
increase of black hole horizon.}\label{fig2}
\end{figure}

In Table.~\ref{table1}(see appendix), we further list the QNM frequencies of massless scalar
perturbation around small and large black holes for a first order SBH/LBH phase transition.
Fixing the pressure with $P=0.003$,
we obtain the phase transition temperature $T_*\simeq0.04567$,
$T_*\simeq0.06945$ and $T_*\simeq0.08664$ in the cases of $k=-1$, 0 and 1, respectively,
where the small and large black hole phases can coexist.
With regard to a small black hole phase, the radius of black hole becomes
smaller and smaller when the temperature decreases from the phase
transition temperature $T_*$. In this process the absolute values of the imaginary part
of the QNM frequencies decrease, while the real part frequencies change very little.
On the other hand, when the temperature for the large black hole phase increases
from the phase transition temperature $T_*$, the black hole gets bigger.
The QNM frequencies increase in the real and absolute value of imaginary parts.
Consequently, the massless scalar perturbation
outside the black hole gets more oscillations but it decays faster. These results
are consistent with the overall discussions reported in \cite{Liu:2014gvf,Chabab:2016cem}.
Figure~\ref{fig2} illustrates the QNM frequencies for small and large black hole phases.
Increase in the black hole size is indicated by the arrows.

%%%%%%%%%%%%%%%%%%%%%%%%%%%%%%%%%%%%%%%%%%%%%%%%%%%%%%%%%%%%%%%%%%%%%%%%%%%

\begin{figure*}
\centering
\includegraphics{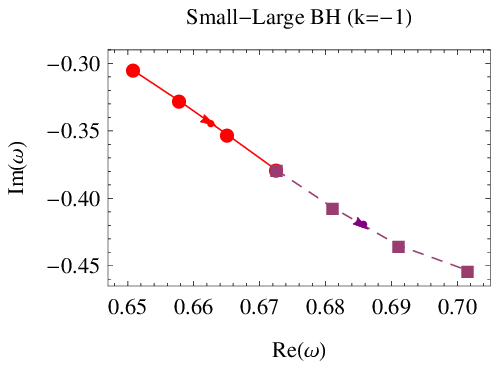}
\hfill%
\includegraphics{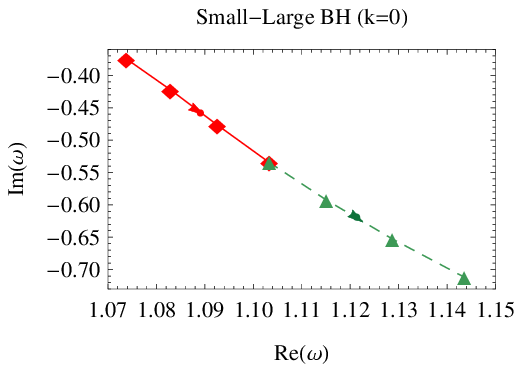}
\hfill%
\includegraphics{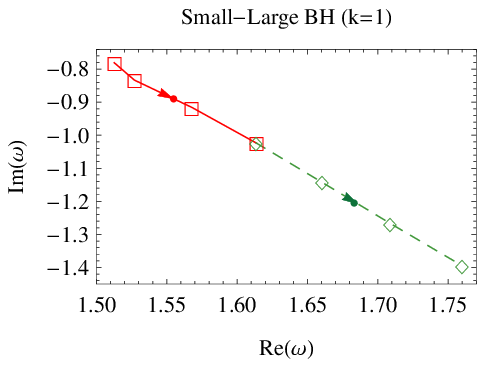}
\caption{The behaviors of QNM frequencies for large and small black holes
in the isobaric process. The arrow indicates the
increase of black hole horizon.}\label{fig3}
\end{figure*}

In addition, at the critical position $P=P_c$, with $P_c\simeq0.0033157$ for $k=-1$,
$P_c\simeq0.0132629$ for $k=0$ and $P_c\simeq0.0298416$ for $k=1$,
a second-order phase transition occurs. The QNM frequencies
of the small and large black hole phases are plotted in Fig.~\ref{fig3}.
We see that QNM frequencies of two black hole phases
show the same behavior as the black hole horizon increases at the critical point.

%%%%%%%%%%%%%%%%%%%%%%%%%%%%%%%%%%%%%%%%%%%%%%%%%%%%%%%%%%%%%%%%%%%%%%%%%%%

\subsection{Isothermal phase transition}

\begin{figure}
\includegraphics[width=2.6in]{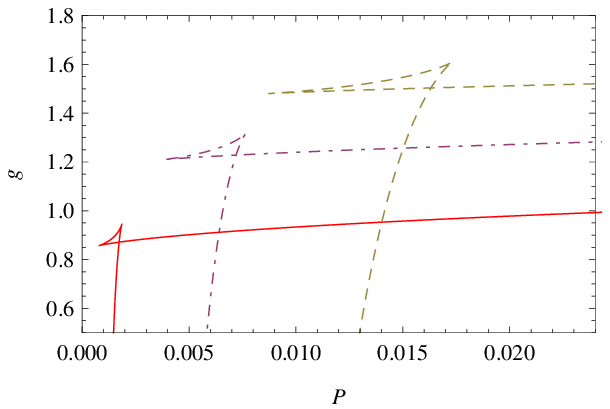}
\hfill%
\includegraphics[width=2.6in]{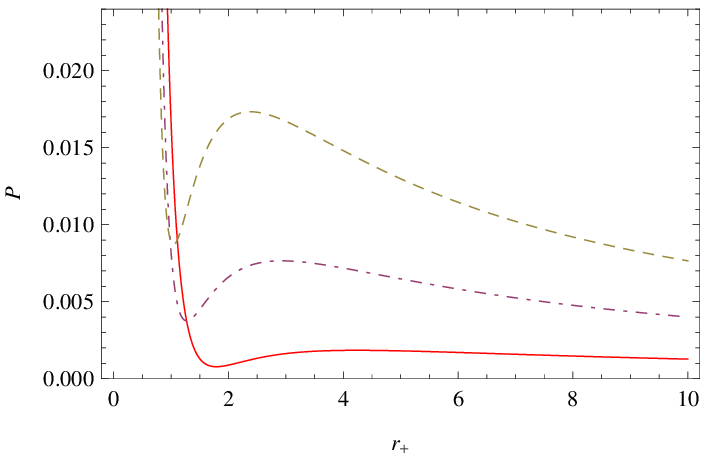}
\caption{The $g-P$(left panel) and $P-r_+$ (right panel) diagrams for $T<T_c$.
The three lines correspond to $k=1$(dashed line), $k=0$(dotdashed line)
and $k=-1$(solid line). }\label{fig4}
\end{figure}

Fixing the black hole temperature $T$, the associated $P-r_+$ diagram
of charged topological AdS black holes is displayed in the right part of Fig.~\ref{fig4}.
For $T<T_c$ there is an inflection point and
behavior is reminiscent of the Van der Waals liquid-gas system. Moreover,
the behavior of Gibbs free energy is plotted in the left panel of
Fig.~\ref{fig4}. Similarly to Fig.~\ref{fig1}, characteristic first order
SBH/LBH phase transition behavior shows up.

\begin{figure}[htb]
\includegraphics[width=2in]{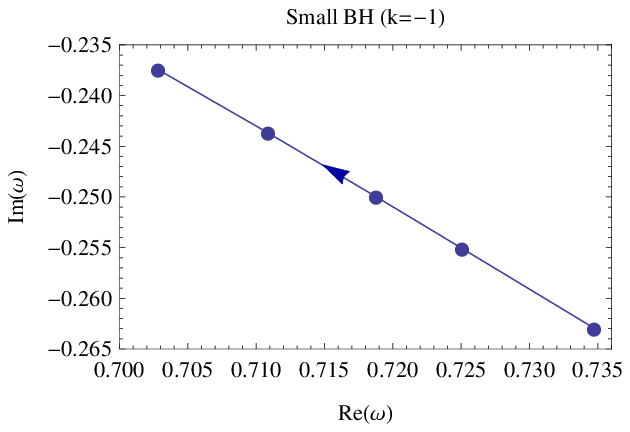}
\hfill%
\includegraphics[width=2in]{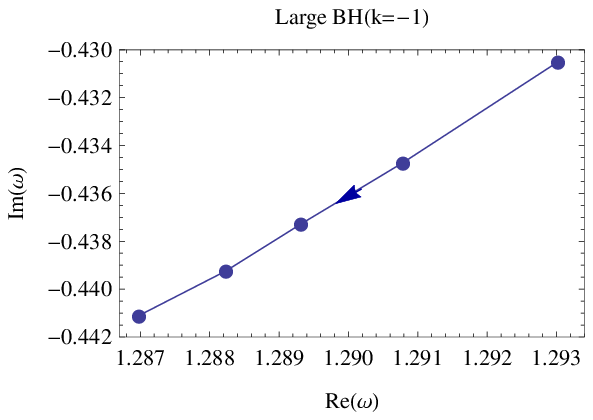}
\hfill%
\includegraphics[width=2in]{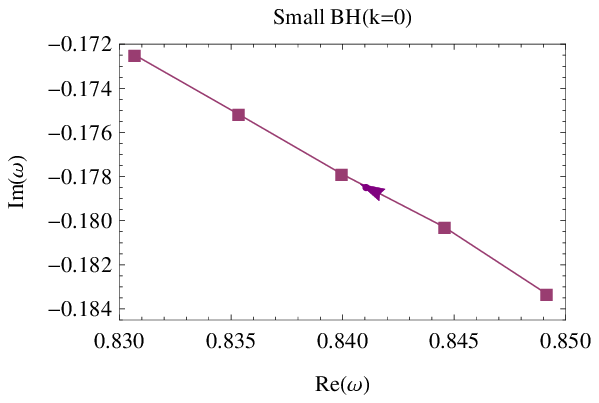}
\hfill%
\includegraphics[width=2in]{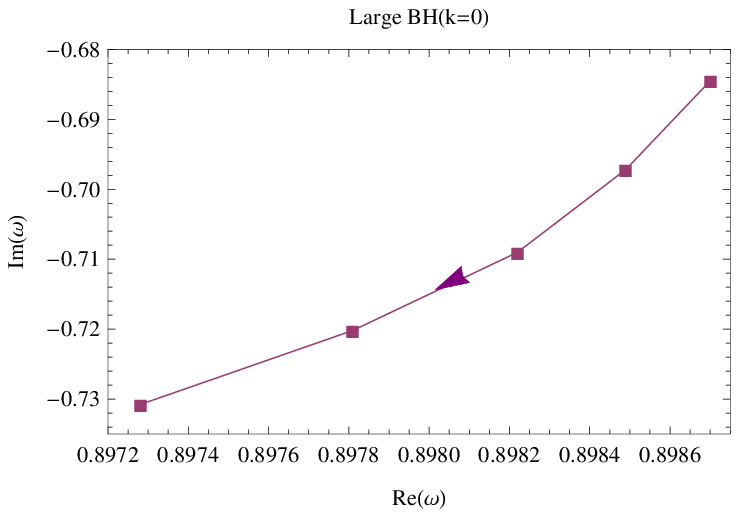}
\hfill%
\includegraphics[width=2in]{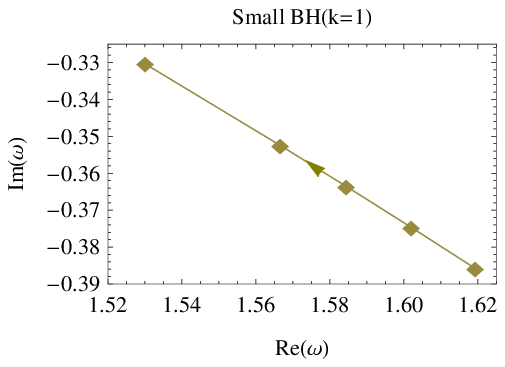}
\hfill%
\includegraphics[width=2in]{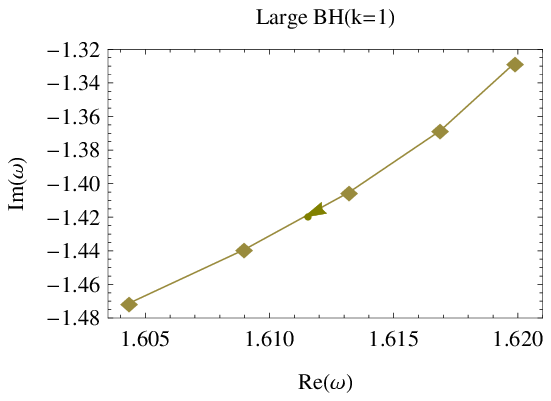}
\caption{The behaviors of QNM frequencies for large and small black holes
in the isothermal process. The arrow indicates the
increase of black hole horizon. }\label{fig5}
\end{figure}

Table~\ref{table2} (see appendix) displays the QNM frequencies of
small and large black hole phases at temperature $T=0.79T_c$
for different horizon topologies $(k=0,\pm1)$ in the isothermal precess.
Then the first order SBH/LBH phase transition happens
at $P_*\approx0.001728$ for $k=-1$,
at $P_*\approx0.007189$ for $k=0$ and at $P_*\approx0.016215$ for $k=1$,
where the small and large black holes possess the same Gibbs free energy
and same pressure. The data above (below) the horizontal line are for the small (large)
black hole phase, respectively. The drastically different QNM
frequencies for small and large black hole phases are plotted in Fig.~\ref{fig5}.
From the figure we see different slopes of the QNM frequencies in
the massless scalar perturbations revealing that small and large black holes
are in different phases.

%%%%%%%%%%%%%%%%%%%%%%%%%%%%%%%%%%%%%%%%%%%%%%%%%%%%%%%%%%%%%%%%%%%%%%%%%%%

%%%%%%%%%%%%%%%%%%%%%%%%%%%%%%%%%%%%%%%%%%%%%%%%%%%%%%%%%%%%%%%%%%%%%%%%%%%

In the isothermal transition, the QNMs can be affected by the value of the
pressure $P(l)$ and the horizon radius $r_+$, which are related by a fixed temperature.
To illustrate the effects of the two parameters, we list the influence of $r_+$ on the
frequencies for small and large black holes by fixing $P(l)$ in Table~\ref{table3} (see appendix)
and QNM frequencies by fixing the black hole size $r_+$ in Table~\ref{table4} (see appendix).
From Tables~\ref{table3} and \ref{table4} one can see that there is competition
between the pressure $P$ and horizon radius $r_+$. Each of these parameters aims
to overwhelm the other which affects the decay rate of the field.

In order to further discuss how these two factors affect the QNM frequencies,
we perform a double-series expansion of the frequency $\omega(r_{+}+\Delta r_+,P+\Delta P)$
\begin{eqnarray}
\omega(r_{+}+\Delta r_+,P+\Delta P)&=&\omega(r_+,P)+\frac{\partial\omega}{\partial r_+}\Delta r_{+}
+\frac{\partial\omega}{\partial P}\Delta P\nonumber\\
&&+\mathcal{O}(\Delta r_+^2,\Delta P^2,\Delta r_{+}\cdot\Delta P).
\end{eqnarray}
Obviously, the changes of the QNM frequency are under two influences, one is from
the change of the black hole size $r_+$ and the other is from the change of the pressure $P$
(or AdS radius $l$). For simple discussions, we define
$\Delta_1\equiv\frac{\partial\omega}{\partial r_+}\Delta r_{+}$ and
$\Delta_2\equiv\frac{\partial\omega}{\partial P}\Delta P$.

Note that the choice of the step of pressure $\Delta P$ in linear approximation is
related to $\Delta r_+$, which is brought about by
\begin{eqnarray}
dP=\left(-\frac{T}{2r_+^2}+\frac{c_{1}m^2}{8\pi r_+}+\frac{k+c_{2}m^2}{4\pi r_+^3}
-\frac{q^2}{8\pi r_+^5}\right)dr_+\label{deos}
\end{eqnarray}
from the equation of state (\ref{eos}). In Table~\ref{table5} (see appendix), we list
the QNM frequencies from the linear approximation for small and large black hole
phase. One can see that the behavior of $\tilde{\omega}$ is in good agreement
with the numerical computation results listed in Table~\ref{table2}.
Comparing $\Delta_1$ and $\Delta_2$ in Table~\ref{table5}, the change of $P$ (or $l$)
in small black hole phase clearly wins over the change of the black hole size,
which dominantly contributes to the behavior of QNM frequencies for small black
hole phase. For the large black hole phase,
the contributions of $\Delta_1$ and $\Delta_2$ on the real part of the QNM
frequency are comparable. But the change of $P$ (or $l$) wins out a little.

\begin{figure*}
\includegraphics[width=2in]{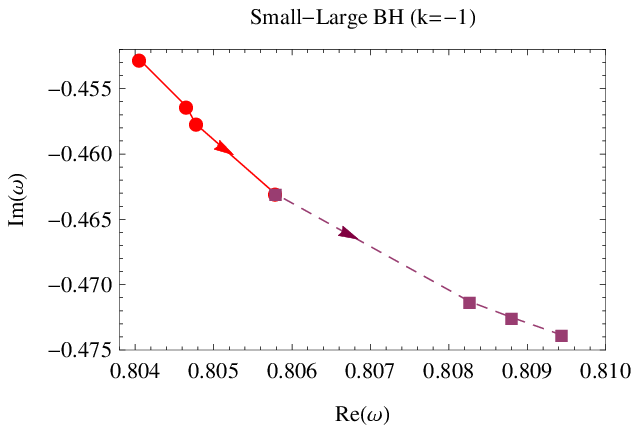}
\hfill%
\includegraphics[width=2in]{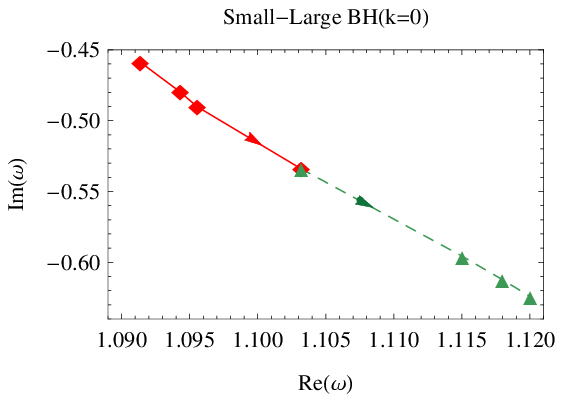}
\hfill%
\includegraphics[width=2in]{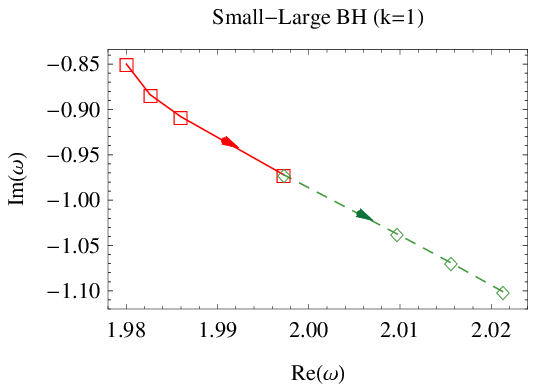}
\caption{The behaviors of QNM frequencies for large and small black holes
in the isothermal process. The arrow indicates the
increase of black hole horizon. }\label{fig6}
\end{figure*}

In addition, for the isothermal phase transition at $T=T_c$, the QNM frequencies
for the small black hole and large black hole are plotted in Fig.~\ref{fig6},
which shows the same behavior as the horizon radius increases.

%%%%%%%%%%%%%%%%%%%%%%%%%%%%%%%%%%%%%%%%%%%%%%%%%%%%%%%%%%%%%%%%%%%%%%%%%%%

\section{Conclusions and discussions}
\label{4s}

We have calculated the QNMs of massless scalar field perturbation around small
and large charged topological AdS black holes in four-dimensional dRGT massive gravity.
When the Van der Waals-like SBH/LBH phase transition
happens in the extended space, no matter whether in the isobaric process by
fixing the pressure $P$ or in the isothermal process by
fixing the temperature $T$ of the system, the slopes of the QNM frequencies
change drastically being different in the small and large black hole phases
as the horizon radius $r_+$ is increasing. This
clearly shows the signature of the phase transition
between small and large black holes. Moreover, we have also found that,
at the critical isothermal and isobaric phase transitions,
QNM frequencies for both small and large black holes have the
same behavior, suggesting that QNMs are not appropriate to
probing the black hole second order phase transition.

Comparing with the action of Eq.~(\ref{action}), Ref.~\cite{Hendi:2017fxp} recently asserted
the existence of a Van der Waals-like SBH/LBH phase transition with
the massive potential ${\cal U}_{3}\neq0$ in the five dimensional case.
Moreover, the charged black hole \cite{Hendi:2015pda},
the Born-Infeld black hole \cite{Hendi:2016yof} and black hole in the
Maxwell and Yang-Mills fields \cite{Meng:2016its}
have recently been  constructed in Gauss-Bonnet massive gravity.
The Van der Waals-like SBH/LBH phase transition also appears in these models.
It would be interesting to extend our discussion to these
black hole solutions.

\begin{acknowledgements}
The work is supported by the National Natural Science Foundation
of China (NNSFC) (Grant No.11605152), and Natural Science Foundation
of Jiangsu Province (Grant No.BK20160452).
D.C.Z. are extremely grateful to
Hai-Qing Zhang and Hua-Bi Zeng for useful discussions.
\end{acknowledgements}

\section{Appendix section}\label{app}

Here we present the related QNM frequencies of the massless scalar perturbation
around small and large black holes in the isobaric as well as in the
isothermal process.

\begin{table*}
\caption{The QNM frequencies of massless scalar perturbation with the
change of black hole temperature in the \textbf{isobaric process}. The upper part,
above the horizontal line, is for the small black hole phase,
while the lower part is for the large black hole phase.}
\label{table1}
\resizebox{17cm}{!}{\begin{tabular*}{\textwidth}{@{\extracolsep{\fill}}lrrrrrrrl@{}}
\hline
\multicolumn{3}{|c|}{$k=-1$}& \multicolumn{3}{|c|}{$k=0$}& \multicolumn{3}{|c|}{$k=1$}\\ \hline
$T(10^{-2})$ & $r_+$ & $\omega$
& $T(10^{-2})$ & $r_+$ & $\omega$
& $T(10^{-2})$ & $r_+$ & $\omega$ \\ \hline
4.3 & 1.53312 & 0.37358-0.08479I & 6.6& 0.84698 &0.58842-0.04675I& 8.3& 0.64697 &0.76024-0.02047I\\
4.35 & 1.56832 & 0.37263-0.08778I& 6.65& 0.84901&0.58840-0.04680I & 8.4& 0.64822&0.76022-0.02049I\\
4.4 &1.61023 & 0.37191-0.09095I & 6.7& 0.85107 &0.58837-0.04684I& 8.45& 0.64886 &0.76021-0.02049I\\
4.45 &1.66230& 0.37132-0.09345I& 6.8& 0.85527&0.58831-0.04694I& 8.5& 0.64950 &0.76020-0.02050I\\
4.5 &1.73196& 0.37053-0.09795I & 6.9& 0.85959 &0.58825-0.04704I& 8.6& 0.65078 &0.76018-0.02051I\\ \hline
4.65& 3.93186 &0.71210-0.47810I & 7.0& 7.49664 &0.62800-0.51924I& 8.7&9.78451 &0.77701-0.67319I\\
4.7 &4.17687 &0.72185-0.49189I & 7.1& 7.79546 &0.63601-0.53979I& 8.75&9.92469 &0.78088-0.68279I \\
4.8 &4.57798& 0.73905-0.51413I & 7.2& 8.0785 &0.64384-0.55920I& 8.8& 10.0623 &0.78472-0.69221I\\
4.85 &4.75357& 0.74717-0.52388I &7.3& 8.34929 &0.65153-0.57774I& 8.9& 10.3307 &0.79235-0.71056I\\
4.9 &4.91841& 0.75514-0.53308I &7.35& 8.48087 &0.65534-0.58673I&9.0& 10.5913 &0.79990-0.72836I\\ \hline
\end{tabular*}}
\end{table*}

\begin{table*}
\caption{The QNM frequencies of massless scalar perturbation with the
change of black hole pressure in the \textbf{isothermal process}.
The upper part, above the horizontal line, is for the
small black hole phase, while the lower part is for the large black hole phase.}
\label{table2}
\resizebox{17cm}{!}{\begin{tabular*}{\textwidth}{@{\extracolsep{\fill}}lrrrrrrrl@{}}
\hline
\multicolumn{3}{|c|}{$k=-1$}&
\multicolumn{3}{|c|}{$k=0$}&
\multicolumn{3}{|c|}{$k=1$}\\  \hline
$P(10^{-3})$ & $r_+$ & $\omega$ &
$P(10^{-3})$& $r_+$ & $\omega$ &
$P(10^{-3})$& $r_+$ & $\omega$ \\ \hline
2.0 & 1.39769 &0.734739-0.262958I & 7.6 &1.01374& 0.84912-0.18330I& 19.5 &0.81148& 1.61923-0.38573I\\
1.95 & 1.40432&0.725094-0.255107I & 7.5 &1.01611& 0.84456-0.18029I& 19.0 &0.81513& 1.60191-0.37460I\\
1.9 &1.40889& 0.718765-0.249971I& 7.4 &1.01853& 0.83996-0.17788I&18.5 &0.81892& 1.58434-0.36350I\\
1.85 &1.41479& 0.710838-0.243658I & 7.3 &1.02101& 0.83533-0.17517I& 18.0 &0.82286& 1.56649-0.35241I\\
1.8 & 1.42092 & 0.702818-0.237435I& 7.2 &1.02354& 0.83065-0.17247I& 17.0 &0.83129& 1.52994-0.33032I \\ \hline
1.6 & 6.90636 & 1.29303-0.430498I& 7.0 &4.26840& 0.89870-0.68450I & 15.0 &3.90321& 1.61987-1.32816I\\
1.46 & 8.13084 & 1.29029-0.434711I & 6.9 &4.41023& 0.89849-0.69717I& 14.5 &4.15993& 1.61688-1.36840I \\
1.4 &8.69728 & 1.28932-0.437272I & 6.8 &4.55105& 0.89822-0.70902I& 14.0 &4.42431& 1.61320-1.40518I \\
1.35 &9.19643  & 1.28823-0.439248I & 6.7 &4.69174& 0.89781-0.72019I& 13.5 &4.69961& 1.60899-1.43925I \\
1.3 &9.72493 &1.28698-0.441104I & 6.6 &4.83294& 0.89728-0.73079I& 13.0 &4.98877& 1.60436-1.47113I \\ \hline
\end{tabular*}}
\end{table*}

\begin{table*}
\caption{The QNM frequencies change as the black hole horizons $r_+$.}
\label{table3}
\resizebox{17cm}{!}{\begin{tabular*}{\textwidth}{@{\extracolsep{\fill}}lrrrrrrrl@{}}
\hline
\multicolumn{3}{|c|}{$k=-1$}&
\multicolumn{3}{|c|}{$k=0$}&
\multicolumn{3}{|c|}{$k=1$}\\  \hline
$P(10^{-3})$ & $r_+$ & $\omega$ &
$P(10^{-3})$& $r_+$ & $\omega$ &
$P(10^{-3})$& $r_+$ & $\omega$ \\ \hline
1.9& 1.32 & 0.719628-0.243002I & 7.4 &1.0& 0.84000-0.17600I& 19.0 &0.7& 1.603743-0.35228I\\
1.9 &1.40889& 0.718765-0.249971I & 7.4 &1.01853& 0.83996-0.17788I& 19.0 &0.815127& 1.601909-0.37460I\\
1.9 & 1.42 &0.718721-0.250594I  & 7.4 &1.2& 0.83982-0.20044I&19.0 &0.95& 1.600035-0.39805I \\ \hline
1.4 &8.6& 1.28319-0.434621I & 6.7 &4.5& 0.88740-0.69084I& 14.0 &4.3& 1.59778-1.36578I \\
1.4 &8.69728 &1.28932-0.437272I & 6.7 &4.69174& 0.89781-0.72019I& 14.0 &4.42431& 1.61320-1.40518I \\
1.4 & 8.8 &1.29586-0.440089I & 6.7 &4.75& 0.90106-0.72910I& 14.0 &4.6& 1.63562-1.46078I \\ \hline
\end{tabular*}}
\end{table*}

\begin{table*}
\caption{The QNM frequencies change as the pressure $P$. }
\label{table4}
\resizebox{17cm}{!}{\begin{tabular*}{\textwidth}{@{\extracolsep{\fill}}lrrrrrrrl@{}}
\hline
\multicolumn{3}{|c|}{$k=-1$}&
\multicolumn{3}{|c|}{$k=0$}&
\multicolumn{3}{|c|}{$k=1$}\\  \hline
$r_+$ & $P(10^{-3})$ & $\omega$ &
$r_+$ & $P(10^{-3})$& $\omega$ &
$r_+$ & $P(10^{-3})$& $\omega$ \\ \hline
1.40889& 1.85 & 0.710848-0.243315I & 1.01853 &7.3& 0.835334-0.1749168I & 0.81513 &18.0& 1.56650-0.35060I\\
1.40889&1.9 & 0.718765-0.24997I & 1.01853& 7.4 &0.8399608-0.177878I & 0.81513& 19.0 &1.60191-0.37460I\\
1.40889 & 2.0 &0.734378-0.263568I &1.01853 &7.5& 0.844556-0.180842I &0.81513&20.0& 1.63634-0.39865I \\ \hline
8.69728 &1.3& 1.22583-0.414773I & 4.69174 &6.6& 0.889526-0.709527I & 4.42431 &13.0& 1.53729-1.30516I \\
8.69728 & 1.4 &1.28932-0.437272I & 4.69174& 6.7 &0.897811-0.720194I & 4.42431& 14.0 &1.61320-1.40518I \\
8.69728 & 1.5 &1.36017-0.458125I & 4.69174 &6.8& 0.906075-0.730857I & 4.42431 &15.0& 1.68843-1.50510I \\ \hline
\end{tabular*}}
\end{table*}

\begin{table*}
\caption{ $\tilde{\omega}$ is the QNM frequencies from the linear approximation.
$\Delta_1$ and $\Delta_2$ represent corrections due to variation of the black
hole size and pressure, respectively. }
\label{table5}
\resizebox{17cm}{!}{\begin{tabular*}{\textwidth}{@{\extracolsep{\fill}}lrrrrl@{}}
\hline
\multicolumn{6}{|c|}{$k=-1$}\\  \hline
$P(10^{-3})$& $r_+$ & $\omega$ &$\tilde{\omega}$ & $\Delta_1$& $\Delta_2$ \\ \hline
2.0 & 1.39769 &0.734739-0.262958I &0.734554-0.262623I& 0.000101584+0.000850I&0.0156869-0.013502I\\ \hline
1.9 &1.40889& 0.718765-0.249971I &0.718765-0.249971I & 0 &0\\ \hline
1.8 & 1.42092 & 0.702818-0.237435I &0.702969-0.237382I &-0.000109-0.000913378I &-0.0156869+0.013502I\\ \hline\hline
1.46 & 8.13084 & 1.29029-0.434711I &1.29375-0.434791I&-0.0358755+0.0154865I &0.0403032-0.0130058I\\ \hline
1.4 &8.69728 &1.28932-0.437272I &1.28932-0.437272I& 0 &0 \\ \hline
1.3 &9.72493 &1.28698-0.441104I  &1.28723-0.443692I &0.0650862-0.028096I &-0.067172+0.0216762I \\ \hline
\hline
\multicolumn{6}{|c|}{$k=0$}\\  \hline
$P(10^{-3})$& $r_+$ &$\omega$ & $\tilde{\omega}$ & $\Delta_1$& $\Delta_2$ \\ \hline
7.5 &1.01611 & 0.84456-0.18029I&0.84457-0.18054I& 0+0.000295724I &0.004611-0.0029626I\\ \hline
7.4 &1.01853& 0.83996-0.17788I&0.83996-0.17788I & 0 &0\\ \hline
7.3 &1.02101&0.83533-0.17517I&0.83534-0.17522I & 0-0.000303056I &-0.004611+0.0029626I\\ \hline\hline
6.8 &4.55105 &0.89822-0.70902I& 0.89839-0.70937I & -0.0076873+0.0214805I &0.008274-0.010665I \\ \hline
6.7 &4.69174& 0.89781-0.72019I&0.89781-0.72019I & 0 &0 \\ \hline
6.6 &4.83294& 0.89728-0.73079I&0.89724-0.73108I &0.00771-0.021558I &-0.008274+0.010665I \\ \hline
\hline
\multicolumn{6}{|c|}{$k=1$}\\  \hline
$P(10^{-3})$& $r_+$ & $\omega$ & $\tilde{\omega}$ & $\Delta_1$& $\Delta_2$ \\ \hline
19.5 &0.81148 & 1.61923-0.38573I &1.61942-0.38594I& 0.000054+0.000668I &0.01746-0.012013I\\ \hline
19.0 &0.81513& 1.60191-0.37460I &1.60191-0.37460I & 0 &0\\ \hline
18.5 &0.81892&1.58434-0.36350I &1.58439-0.36198I & -0.000056-0.0006I &-0.01746+0.01202I\\ \hline\hline
14.5 &4.15993 &1.61688-1.36840I &1.61763-1.37144I & -0.033347+0.08372I &0.037785-0.049985I \\ \hline
14.0 &4.42431&1.61320-1.40518I & 1.61320-1.40518I & 0 &0 \\ \hline
13.5 &4.69961& 1.60899-1.43925I &1.61014-1.44237I &0.034725-0.087178I &-0.037785+0.049985I \\ \hline
\end{tabular*}}
\end{table*}


\begin{thebibliography}{9}

%\cite{Gupta:1954zz}
\bibitem{Gupta:1954zz}
  S.~N.~Gupta,
%  ``Gravitation and Electromagnetism,''
  Phys. Rev. \textbf{96}, 1683 (1954).
  %%CITATION = PHRVA,96,1683;%%
  %109 citations counted in INSPIRE as of 11 Mar 2015
%\cite{Weinberg:1965rz}
\bibitem{Weinberg:1965rz}
  S.~Weinberg,
 % ``Photons and gravitons in perturbation theory: Derivation of Maxwell's and Einstein's equations,''
  Phys. Rev. \textbf{138}, B988 (1965).
  %%CITATION = PHRVA,138,B988;%%
  %257 citations counted in INSPIRE as of 11 mar 2015
%\cite{Feynman:1996kb}
\bibitem{Feynman:1996kb}
  R.~P.~Feynman, F.~B.~Morinigo, W.~G.~Wagner and B.~Hatfield,
%  ``Feynman lectures on gravitation,''
  Reading, USA: Addison-Wesley (1995) 232 p. (The advanced book program)
  %4 citations counted in INSPIRE as of 11 mar 2015
%\cite{Hassan:2011zd}
\bibitem{Hassan:2011zd}
  S.~F.~Hassan and R.~A.~Rosen,
  %``Bimetric Gravity from Ghost-free Massive Gravity,''
  JHEP \textbf{1202}, 126 (2012)
  [arXiv:1109.3515 [hep-th]].
  %%CITATION = doi:10.1007/JHEP02(2012)126;%%
  %396 citations counted in INSPIRE as of 28 Mar 2017
 %\cite{DAmico:2011eto}
\bibitem{DAmico:2011eto}
  G.~D'Amico, C.~de Rham, S.~Dubovsky, G.~Gabadadze, D.~Pirtskhalava and A.~J.~Tolley,
  %``Massive Cosmologies,''
  Phys. Rev. D \textbf{84}, 124046 (2011)
  [arXiv:1108.5231 [hep-th]].
  %%CITATION = doi:10.1103/PhysRevD.84.124046;%%
  %261 citations counted in INSPIRE as of 28 Mar 2017
  %\cite{Akrami:2012vf}
\bibitem{Akrami:2012vf}
  Y.~Akrami, T.~S.~Koivisto and M.~Sandstad,
  %``Accelerated expansion from ghost-free bigravity: a statistical analysis with improved generality,''
  JHEP \textbf{1303}, 099 (2013)
  [arXiv:1209.0457 [astro-ph.CO]].
  %%CITATION = doi:10.1007/JHEP03(2013)099;%%
  %115 citations counted in INSPIRE as of 28 Mar 2017
%\cite{Akrami:2015qga}
\bibitem{Akrami:2015qga}
  Y.~Akrami, S.~F.~Hassan, F.~Könnig, A.~Schmidt-May and A.~R.~Solomon,
  %``Bimetric gravity is cosmologically viable,''
  Phys. Lett. B \textbf{748}, 37 (2015)
  [arXiv:1503.07521 [gr-qc]].
  %%CITATION = doi:10.1016/j.physletb.2015.06.062;%%
  %45 citations counted in INSPIRE as of 28 Mar 2017
%\cite{Abbott:2016blz}
\bibitem{Abbott:2016blz}
  B.~P.~Abbott {\it et al.} [LIGO Scientific and Virgo Collaborations],
  %``Observation of Gravitational Waves from a Binary Black Hole Merger,''
  Phys. Rev. Lett.  \textbf{116}, 061102 (2016)
  [arXiv:1602.03837 [gr-qc]].
  %%CITATION = doi:10.1103/PhysRevLett.116.061102;%%
  %1237 citations counted in INSPIRE as of 28 Mar 2017
%\cite{Fierz:1939ix}
\bibitem{Fierz:1939ix}
  M.~Fierz and W.~Pauli,
  %``On relativistic wave equations for particles of arbitrary spin in an electromagnetic field,''
  Proc. Roy. Soc. Lond. A \textbf{173}, 211 (1939).
  %%CITATION = doi:10.1098/rspa.1939.0140;%%
  %1054 citations counted in INSPIRE as of 09 Feb 2017
%\cite{Boulware:1973my}
\bibitem{Boulware:1973my}
  D.~G.~Boulware and S.~Deser,
%  ``Can gravitation have a finite range?,''
  Phys. Rev. D \textbf{6}, 3368 (1972).
  %%CITATION = PHRVA,D6,3368;%%
  %529 citations counted in INSPIRE as of 11 Mar 2015
%\cite{deRham:2010ik}
\bibitem{deRham:2010ik}
  C.~de Rham and G.~Gabadadze,
%  ``Generalization of the Fierz-Pauli Action,''
  Phys. Rev. D \textbf{82}, 044020 (2010)
  [arXiv:1007.0443 [hep-th]].
  %%CITATION = ARXIV:1007.0443;%%
  %429 citations counted in INSPIRE as of 11 Mar 2015\cite{deRham:2010ik,deRham:2010kj}
%\cite{deRham:2010kj}
\bibitem{deRham:2010kj}
  C.~de Rham, G.~Gabadadze and A.~J.~Tolley,
%  ``Resummation of Massive Gravity,''
  Phys. Rev. Lett. \textbf{106}, 231101 (2011)
  [arXiv:1011.1232 [hep-th]].
  %%CITATION = ARXIV:1011.1232;%%
  %528 citations counted in INSPIRE as of 11 Mar 2015
  %\cite{Vegh:2013sk}
\bibitem{Vegh:2013sk}
  D.~Vegh,
%  ``Holography without translational symmetry,''
  arXiv:1301.0537 [hep-th].
  %%CITATION = ARXIV:1301.0537;%%
  %61 citations counted in INSPIRE as of 11 Mar 2015
%\cite{Ghosh:2015cva}
\bibitem{Ghosh:2015cva}
  S.~G.~Ghosh, L.~Tannukij and P.~Wongjun,
  %``A class of black holes in dRGT massive gravity and their thermodynamical properties,''
  Eur. Phys. J. C \textbf{76}, 119 (2016)
  [arXiv:1506.07119 [gr-qc]].
  %%CITATION = doi:10.1140/epjc/s10052-016-3943-x;%%
  %13 citations counted in INSPIRE as of 09 Feb 2017
%\cite{Do:2016abo}
\bibitem{Do:2016abo}
  T.~Q.~Do,
  %``Higher dimensional nonlinear massive gravity,''
  Phys. Rev. D \textbf{93}, 104003 (2016) [arXiv:1602.05672 [gr-qc]].
  %%CITATION = doi:10.1103/PhysRevD.93.104003;%%  %6 citations counted in INSPIRE as of 23 Feb 2017
%\cite{Li:2016fbf}
\bibitem{Li:2016fbf}
  P.~Li, X.~z.~Li and P.~Xi,
  %``Black hole solutions in de Rham-Gabadadze-Tolley massive gravity,''
  Phys. Rev. D \textbf{93}, 064040 (2016)
  [arXiv:1603.06039 [gr-qc]].
  %%CITATION = doi:10.1103/PhysRevD.93.064040;%%
  %9 citations counted in INSPIRE as of 09 Feb 2017
%\cite{Cai:2012db}
\bibitem{Cai:2012db}
  Y.~F.~Cai, D.~A.~Easson, C.~Gao and E.~N.~Saridakis,
  %``Charged black holes in nonlinear massive gravity,''
  Phys. Rev. D \textbf{87}, 064001 (2013) [arXiv:1211.0563 [hep-th]].
  %%CITATION = doi:10.1103/PhysRevD.87.064001;%%  %38 citations counted in INSPIRE as of 23 Feb 2017
%\cite{Berezhiani:2011mt}
\bibitem{Berezhiani:2011mt}
  L.~Berezhiani, G.~Chkareuli, C.~de Rham, G.~Gabadadze and A.~J.~Tolley,
  %``On Black Holes in Massive Gravity,''
  Phys. Rev. D \textbf{85}, 044024 (2012) [arXiv:1111.3613 [hep-th]].
  %%CITATION = doi:10.1103/PhysRevD.85.044024;%%  %112 citations counted in INSPIRE as of 23 Feb 2017
%\cite{Hendi:2015hoa}
\bibitem{Hendi:2015hoa}
  S.~H.~Hendi, B.~Eslam Panah and S.~Panahiyan,
  %``Einstein-Born-Infeld-Massive Gravity: adS-Black Hole
  %Solutions and their Thermodynamical properties,''
  JHEP \textbf{1511}, 157 (2015)
  [arXiv:1508.01311 [hep-th]].
  %%CITATION = doi:10.1007/JHEP11(2015)157;%%
  %29 citations counted in INSPIRE as of 09 Feb 2017

 %\cite{Tannukij:2017jtn}
\bibitem{Tannukij:2017jtn}
  L.~Tannukij, P.~Wongjun and S.~G.~Ghosh,
  %``Black String in dRGT Massive Gravity,''
  arXiv:1701.05332 [gr-qc].
  %%CITATION = ARXIV:1701.05332;%%
%\cite{Hendi:2016hbe}
\bibitem{Hendi:2016hbe}
  S.~H.~Hendi, S.~Panahiyan, S.~Upadhyay and B.~Eslam Panah,
  %``Charged BTZ black holes in the context of massive gravity's rainbow,''
  Phys.\ Rev.\ D \textbf{95}, no. 8, 084036 (2017)
  [arXiv:1611.02937 [hep-th]].
  %%CITATION = doi:10.1103/PhysRevD.95.084036;%%
  %10 citations counted in INSPIRE as of 30 Apr 2017
%\cite{Hendi:2016pvx}
\bibitem{Hendi:2016pvx}
  S.~H.~Hendi, B.~Eslam Panah and S.~Panahiyan,
  %``Massive charged BTZ black holes in asymptotically (a)dS spacetimes,''
  JHEP \textbf{1605}, 029 (2016)
  [arXiv:1604.00370 [hep-th]].
  %%CITATION = doi:10.1007/JHEP05(2016)029;%%
  %17 citations counted in INSPIRE as of 09 Feb 2017
%\cite{Hendi:2015pda}
\bibitem{Hendi:2015pda}
  S.~H.~Hendi, S.~Panahiyan and B.~Eslam Panah,
  %``Charged Black Hole Solutions in Gauss-Bonnet-Massive Gravity,''
  JHEP \textbf{1601}, 129 (2016)
  [arXiv:1507.06563 [hep-th]].
  %%CITATION = doi:10.1007/JHEP01(2016)129;%%
  %30 citations counted in INSPIRE as of 09 Feb 2017

%\cite{Meng:2016its}
\bibitem{Meng:2016its}
  K.~Meng and J.~Li,
  %``Black hole solution of Gauss-Bonnet massive gravity coupled to Maxwell and Yang-Mills fields in five dimensions,''
  Europhys. Lett.  \textbf{116},  10005 (2016).
  %%CITATION = doi:10.1209/0295-5075/116/10005;%%  %1 citations counted in INSPIRE as of 23 Feb 2017
%\cite{Zeng:2014uoa}
\bibitem{Zeng:2014uoa}
  H.~B.~Zeng and J.~P.~Wu,
  %``Holographic superconductors from the massive gravity,''
  Phys. Rev. D \textbf{90},  046001 (2014)  [arXiv:1404.5321 [hep-th]].
  %%CITATION = doi:10.1103/PhysRevD.90.046001;%%  %23 citations counted in INSPIRE as of 23 Feb 2017
%\cite{Hendi:2016uni}
\bibitem{Hendi:2016uni}
  S.~H.~Hendi, N.~Riazi and S.~Panahiyan,
  %``Holographical aspects of dyonic black holes: Massive gravity generalization,''
  arXiv:1610.01505 [hep-th].
  %%CITATION = ARXIV:1610.01505;%%
  %4 citations counted in INSPIRE as of 21 Feb 2017
%\cite{Ge:2014aza}
\bibitem{Ge:2014aza}
  X.~H.~Ge, Y.~Ling, C.~Niu and S.~J.~Sin,
  %``Thermoelectric conductivities, shear viscosity, and stability in an anisotropic linear axion model,''
  Phys. Rev. D \textbf{92},  106005 (2015)  [arXiv:1412.8346 [hep-th]].
  %%CITATION = doi:10.1103/PhysRevD.92.106005;%%  %36 citations counted in INSPIRE as of 23 Feb 2017
%\cite{Baggioli:2014roa}
\bibitem{Baggioli:2014roa}
  M.~Baggioli and O.~Pujolas,
  %``Electron-Phonon Interactions, Metal-Insulator Transitions, and Holographic Massive Gravity,''
  Phys. Rev. Lett.  \textbf{114},  251602 (2015)  [arXiv:1411.1003 [hep-th]].
  %%CITATION = doi:10.1103/PhysRevLett.114.251602;%%  %51 citations counted in INSPIRE as of 23 Feb 2017
%\cite{Hu:2015dnl}
\bibitem{Hu:2015dnl}
  Y.~P.~Hu, H.~F.~Li, H.~B.~Zeng and H.~Q.~Zhang,
  %``Holographic Josephson Junction from Massive Gravity,''
  Phys. Rev. D \textbf{93}, 104009 (2016)  [arXiv:1512.07035 [hep-th]].
  %%CITATION = doi:10.1103/PhysRevD.93.104009;%%  %6 citations counted in INSPIRE as of 23 Feb 2017
%\cite{Sadeghi:2015vaa}
\bibitem{Sadeghi:2015vaa}
  M.~Sadeghi and S.~Parvizi,
  %``Hydrodynamics of a black brane in Gauss¨CBonnet massive gravity,''
Class. Quant. Grav. \textbf{33}, 035005 (2016) [arXiv:1507.07183 [hep-th]].
%%CITATION = doi:10.1088/0264-9381/33/3/035005;%%  %7 citations counted in INSPIRE as of 23 Feb 2017
%\cite{Hendi:2015bna}
\bibitem{Hendi:2015bna}
  S.~H.~Hendi, B.~Eslam Panah and S.~Panahiyan,
  %``Thermodynamical Structure of AdS Black Holes in Massive Gravity with Stringy Gauge-Gravity Corrections,''
  Class. Quant. Grav. \textbf{33},  235007 (2016)  [arXiv:1510.00108 [hep-th]].
  %%CITATION = doi:10.1088/0264-9381/33/23/235007;%%  %9 citations counted in INSPIRE as of 23 Feb 2017
  %\cite{Hendi:2016yof}
\bibitem{Hendi:2016yof}
  S.~H.~Hendi, G.~Q.~Li, J.~X.~Mo, S.~Panahiyan and B.~Eslam Panah,
  %``New perspective for black hole thermodynamics in Gauss¨CBonnet¨CBorn¨CInfeld massive gravity,''
  Eur. Phys. J. C \textbf{76},  571 (2016)
  [arXiv:1608.03148 [gr-qc]].
  %%CITATION = doi:10.1140/epjc/s10052-016-4410-4;%%
  %4 citations counted in INSPIRE as of 09 Feb 2017
%\cite{Cai:2014znn}
\bibitem{Cai:2014znn}
  R.~G.~Cai, Y.~P.~Hu, Q.~Y.~Pan and Y.~L.~Zhang,
  %``Thermodynamics of Black Holes in Massive Gravity,''
  Phys. Rev. D \textbf{91}, 024032 (2015)
  [arXiv:1409.2369 [hep-th]].
  %%CITATION = doi:10.1103/PhysRevD.91.024032;%%
  %43 citations counted in INSPIRE as of 09 Feb 2017
  %\cite{Adams:2014vza}
\bibitem{Adams:2014vza}
  A.~Adams, D.~A.~Roberts and O.~Saremi,
 % ``Hawking-Page transition in holographic massive gravity,''
  Phys. Rev. D \textbf{91}, 046003 (2015)
  [arXiv:1408.6560 [hep-th]].
  %%CITATION = ARXIV:1408.6560;%%
  %4 citations counted in INSPIRE as of 11 Mar 2015
%\cite{Prasia:2016esx}
\bibitem{Prasia:2016esx}
  P.~Prasia and V.~C.~Kuriakose,
  %``Quasinormal Modes and Thermodynamics of Linearly Charged BTZ Black holes
  %in Massive Gravity in (Anti)de Sitter Space Time,''
  Eur.\ Phys.\ J.\ C \textbf{77}, no. 1, 27 (2017)
  [arXiv:1608.05299 [gr-qc]].
  %%CITATION = doi:10.1140/epjc/s10052-016-4591-x;%%
  %2 citations counted in INSPIRE as of 09 Feb 2017

%\cite{Caldarelli:1999xj}
\bibitem{Caldarelli:1999xj}
  M.~M.~Caldarelli, G.~Cognola and D.~Klemm,
  %``Thermodynamics of Kerr-Newman-AdS black holes and conformal field theories,''
  Class. Quant. Grav.  \textbf{17}, 399 (2000)
  [hep-th/9908022].
  %%CITATION = doi:10.1088/0264-9381/17/2/310;%%
  %357 citations counted in INSPIRE as of 24 Mar 2017
%\cite{Kastor:2009wy}
\bibitem{Kastor:2009wy}
  D.~Kastor, S.~Ray and J.~Traschen,
  %``Enthalpy and the Mechanics of AdS Black Holes,''
  Class. Quant. Grav.  \textbf{26}, 195011 (2009)
  [arXiv:0904.2765 [hep-th]].
  %%CITATION = doi:10.1088/0264-9381/26/19/195011;%%
  %270 citations counted in INSPIRE as of 24 Mar 2017
 %\cite{Lu:2012xu}
\bibitem{Lu:2012xu}
  H.~Lu, Y.~Pang, C.~N.~Pope and J.~F.~Vazquez-Poritz,
  %``AdS and Lifshitz Black Holes in Conformal and Einstein-Weyl Gravities,''
  Phys. Rev. D \textbf{86}, 044011 (2012)
  [arXiv:1204.1062 [hep-th]].
  %%CITATION = doi:10.1103/PhysRevD.86.044011;%%
  %97 citations counted in INSPIRE as of 24 Mar 2017
%\cite{Kubiznak:2012wp}
\bibitem{Kubiznak:2012wp}
  D.~Kubiznak and R.~B.~Mann,
  %``P-V criticality of charged AdS black holes,''
  JHEP \textbf{1207}, 033 (2012)  [arXiv:1205.0559 [hep-th]].
   %%CITATION = ARXIV:1205.0559;%%  %21 citations counted in INSPIRE as of 09 Oct 2013

%\cite{Gunasekaran:2012dq}
\bibitem{Gunasekaran:2012dq}
  S.~Gunasekaran, R.~B.~Mann and D.~Kubiznak,
  %``Extended phase space thermodynamics for charged and
  %rotating black holes and Born-Infeld vacuum polarization,''
  JHEP \textbf{1211}, 110 (2012)  [arXiv:1208.6251 [hep-th]].
  %%CITATION = ARXIV:1208.6251;%%  %16 citations counted in INSPIRE as of 09 Oct 2013
%\cite{Hendi:2012um}
\bibitem{Hendi:2012um}
  S.~H.~Hendi and M.~H.~Vahidinia,
  %``P-V criticality of higher dimensional black holes with nonlinear source,''
  Phys. Rev. D \textbf{88}, 084045 (2013)  [arXiv:1212.6128 [hep-th]].
  %%CITATION = ARXIV:1212.6128;%%  %5 citations counted in INSPIRE as of 29 Nov 2013
%\cite{Hendi:2015hgg}
\bibitem{Hendi:2015hgg}
  S.~H.~Hendi, R.~M.~Tad, Z.~Armanfard and M.~S.~Talezadeh,
  %``Extended phase space thermodynamics and P–V criticality: Brans–Dicke–Born–Infeld
  %vs. Einstein–Born–Infeld-dilaton black holes,''
  Eur. Phys. J. C \textbf{76}, 263 (2016)
  [arXiv:1511.02761 [gr-qc]].
  %%CITATION = doi:10.1140/epjc/s10052-016-4106-9;%%
  %3 citations counted in INSPIRE as of 23 Mar 2017
%\cite{Zhao:2013oza}
\bibitem{Zhao:2013oza}
  R.~Zhao, H.~-H.~Zhao, M.~-S.~Ma and L.~-C.~Zhang,
  %``On the critical phenomena and thermodynamics of charged topological dilaton AdS black holes,''
  Eur. Phys. J. C \textbf{73}, 2645 (2013)  [arXiv:1305.3725 [gr-qc]].
  %%CITATION = ARXIV:1305.3725;%%  %10 citations counted in INSPIRE as of 12 Mar 2014
%\cite{Zou:2013owa}
\bibitem{Zou:2013owa}
  D.~C.~Zou, S.~J.~Zhang and B.~Wang,
  %``Critical behavior of Born-Infeld AdS black holes in the extended phase space thermodynamics,''
  Phys. Rev. D \textbf{89}, 044002 (2014)  [arXiv:1311.7299 [hep-th]].
  %%CITATION = ARXIV:1311.7299;%%  %4 citations counted in INSPIRE as of 12 Mar 2014
%\cite{Zou:2014mha}
\bibitem{Zou:2014mha}
  D.~C.~Zou, Y.~Liu and B.~Wang,
  %``Critical behavior of charged Gauss-Bonnet AdS black holes in the grand canonical ensemble,''
  Phys. Rev. D \textbf{90}, 044063 (2014)
  [arXiv:1404.5194 [hep-th]].
  %%CITATION = doi:10.1103/PhysRevD.90.044063;%%
  %58 citations counted in INSPIRE as of 10 Feb 2017
%\cite{Cai:2013qga}
\bibitem{Cai:2013qga}
  R.~G.~Cai, L.~M.~Cao, L.~Li and R.~Q.~Yang,
  %``P-V criticality in the extended phase space of Gauss-Bonnet black holes in AdS space,''
  JHEP \textbf{1309}, 005 (2013)  [arXiv:1306.6233 [gr-qc]].
  %%CITATION = ARXIV:1306.6233;%%  %2 citations counted in INSPIRE as of 09 Oct 2013
  %5 citations counted in INSPIRE as of 17 Oct 2013
%\cite{Dehghani:2014caa}
\bibitem{Dehghani:2014caa}
  M.~H.~Dehghani, S.~Kamrani and A.~Sheykhi,
  %``$P-V$ criticality of charged dilatonic black holes,''
  Phys. Rev. D \textbf{90}, 104020 (2014)
  [arXiv:1505.02386 [hep-th]].
  %%CITATION = doi:10.1103/PhysRevD.90.104020;%%
  %17 citations counted in INSPIRE as of 10 Feb 2017
%\cite{Mo:2014qsa}
\bibitem{Mo:2014qsa}
  J.~X.~Mo and W.~B.~Liu,
  %``P-V Criticality of Topological Black Holes in Lovelock-Born-Infeld Gravity,''
  Eur. Phys. J. C \textbf{74}, 2836 (2014)  [arXiv:1401.0785 [gr-qc]].
  %%CITATION = ARXIV:1401.0785;%%  %3 citations counted in INSPIRE as of 20 Apr 2014
%\cite{Hennigar:2015esa}
\bibitem{Hennigar:2015esa}
  R.~A.~Hennigar, W.~G.~Brenna and R.~B.~Mann,
  %``$P-v$ criticality in quasitopological gravity,''
  JHEP \textbf{1507}, 077 (2015)
  [arXiv:1505.05517 [hep-th]].
  %%CITATION = doi:10.1007/JHEP07(2015)077;%%
  %29 citations counted in INSPIRE as of 10 Feb 2017
%\cite{Zhang:2014jfa}
\bibitem{Zhang:2014jfa}
  L.~C.~Zhang, M.~S.~Ma, H.~H.~Zhao and R.~Zhao,
  %``Thermodynamics of phase transition in higher-dimensional Reissner-Nordström-de Sitter black hole,''
  Eur. Phys. J. C \textbf{74}, 3052 (2014)
  [arXiv:1403.2151 [gr-qc]].
  %%CITATION = doi:10.1140/epjc/s10052-014-3052-7;%%
  %22 citations counted in INSPIRE as of 23 Mar 2017
%\cite{Xu:2013zea}
\bibitem{Xu:2013zea}
  W.~Xu, H.~Xu and L.~Zhao,
  %``Gauss-Bonnet coupling constant as a free thermodynamical variable and the associated criticality,''
  Eur. Phys. J. C \textbf{74}, 2970 (2014)
  [arXiv:1311.3053 [gr-qc]].
  %%CITATION = doi:10.1140/epjc/s10052-014-2970-8;%%
  %53 citations counted in INSPIRE as of 23 Mar 2017
 %\cite{Rajagopal:2014ewa}
\bibitem{Rajagopal:2014ewa}
  A.~Rajagopal, D.~Kubiz¨¾¨¢k and R.~B.~Mann,
  %``Van der Waals black hole,''
  Phys. Lett. B \textbf{737}, 277 (2014)
  [arXiv:1408.1105 [gr-qc]].
  %%CITATION = doi:10.1016/j.physletb.2014.08.054;%%
  %23 citations counted in INSPIRE as of 10 Feb 2017
%\cite{Frassino:2014pha}
\bibitem{Frassino:2014pha}
  A.~M.~Frassino, D.~Kubiznak, R.~B.~Mann and F.~Simovic,
  %``Multiple Reentrant Phase Transitions and Triple Points in Lovelock Thermodynamics,''
  JHEP \textbf{1409}, 080 (2014)
  [arXiv:1406.7015 [hep-th]].
  %%CITATION = doi:10.1007/JHEP09(2014)080;%%
  %63 citations counted in INSPIRE as of 10 Feb 2017
%\cite{Wei:2014hba}
\bibitem{Wei:2014hba}
  S.~W.~Wei and Y.~X.~Liu,
  %``Triple points and phase diagrams in the extended phase space of charged Gauss-Bonnet black holes in AdS space,''
  Phys. Rev. D \textbf{90},  044057 (2014)
  [arXiv:1402.2837 [hep-th]].
  %%CITATION = doi:10.1103/PhysRevD.90.044057;%%
  %52 citations counted in INSPIRE as of 10 Feb 2017
%\cite{Altamirano:2014tva}
\bibitem{Altamirano:2014tva}
  N.~Altamirano, D.~Kubiznak, R.~B.~Mann and Z.~Sherkatghanad,
  %``Thermodynamics of rotating black holes and black rings: phase transitions and thermodynamic volume,''
  Galaxies \textbf{2}, 89 (2014)
  [arXiv:1401.2586 [hep-th]].
  %%CITATION = doi:10.3390/galaxies2010089;%%
  %106 citations counted in INSPIRE as of 10 Feb 2017
%\cite{Wei:2015iwa}
\bibitem{Wei:2015iwa}
  S.~W.~Wei and Y.~X.~Liu,
  %``Insight into the Microscopic Structure of an AdS Black Hole from a Thermodynamical Phase Transition,''
  Phys. Rev. Lett.  \textbf{115}, 111302 (2015)
  Erratum: [Phys. Rev. Lett. \textbf{116},  169903 (2016)]
  [arXiv:1502.00386 [gr-qc]].
  %%CITATION = doi:10.1103/PhysRevLett.116.169903, 10.1103/PhysRevLett.115.111302;%%
  %30 citations counted in INSPIRE as of 03 May 2017
  %\cite{Cheng:2016bpx}
\bibitem{Cheng:2016bpx}
  P.~Cheng, S.~W.~Wei and Y.~X.~Liu,
  %``Critical phenomena in the extended phase space of Kerr-Newman-AdS black holes,''
  Phys. Rev. D \textbf{94}, 024025 (2016)
  [arXiv:1603.08694 [gr-qc]].
  %%CITATION = doi:10.1103/PhysRevD.94.024025;%%
  %3 citations counted in INSPIRE as of 03 May 2017
  %\cite{Wei:2015ana}
\bibitem{Wei:2015ana}
  S.~W.~Wei, P.~Cheng and Y.~X.~Liu,
  %``Analytical and exact critical phenomena of $d$-dimensional singly spinning Kerr-AdS black holes,''
  Phys. Rev. D \textbf{93}, 084015 (2016)
  [arXiv:1510.00085 [gr-qc]].
  %%CITATION = doi:10.1103/PhysRevD.93.084015;%%
  %8 citations counted in INSPIRE as of 03 May 2017
%\cite{Hendi:2015cqz}
\bibitem{Hendi:2015cqz}
  S.~H.~Hendi, S.~Panahiyan, B.~E.~Panah and Z.~Armanfard,
  %``Phase transition of charged Black Holes in Brans–Dicke theory through geometrical thermodynamics,''
  Eur. Phys. J. C \textbf{76}, 396 (2016)
  [arXiv:1511.00598 [gr-qc]].
  %%CITATION = doi:10.1140/epjc/s10052-016-4235-1;%%
  %6 citations counted in INSPIRE as of 23 Mar 2017
 %\cite{B.:2015koa}
\bibitem{B.:2015koa}
  C.~B.~Prasobh, J.~Suresh and V.~C.~Kuriakose,
  %``Thermodynamics of Charged Lovelock - AdS Black Holes,''
  Eur. Phys. J. C \textbf{76}, 207 (2016)
  [arXiv:1510.04784 [gr-qc]].
  %%CITATION = doi:10.1140/epjc/s10052-016-4062-4;%%
  %3 citations counted in INSPIRE as of 23 Mar 2017
%\cite{Mo:2016ndm}
\bibitem{Mo:2016ndm}
  J.~X.~Mo, G.~Q.~Li and X.~B.~Xu,
  %``Combined effects of $f(R)$ gravity and conformally invariant Maxwell field on the extended phase space thermodynamics of higher-dimensional black holes,''
  Eur. Phys. J. C \textbf{76},  545 (2016)
  [arXiv:1609.06422 [gr-qc]].
  %%CITATION = doi:10.1140/epjc/s10052-016-4391-3;%%
  %1 citations counted in INSPIRE as of 23 Mar 2017
%\cite{Belhaj:2014eha}
\bibitem{Belhaj:2014eha}
  A.~Belhaj, M.~Chabab, H.~El moumni, K.~Masmar and M.~B.~Sedra,
  %``Maxwell`s equal-area law for Gauss-Bonnet-Anti-de Sitter black holes,''
  Eur. Phys. J. C \textbf{75},  71 (2015)
  [arXiv:1412.2162 [hep-th]].
  %%CITATION = doi:10.1140/epjc/s10052-015-3299-7;%%
  %16 citations counted in INSPIRE as of 23 Mar 2017
%\cite{Xu:2014kwa}
\bibitem{Xu:2014kwa}
  W.~Xu and L.~Zhao,
  %``Critical phenomena of static charged AdS black holes in conformal gravity,''
  Phys. Lett. B \textbf{736}, 214 (2014)
  [arXiv:1405.7665 [gr-qc]].
  %%CITATION = doi:10.1016/j.physletb.2014.07.019;%%
  %38 citations counted in INSPIRE as of 10 Feb 201
%\cite{Xu:2014tja}
\bibitem{Xu:2014tja}
  H.~Xu, W.~Xu and L.~Zhao,
  %``Extended phase space thermodynamics for third order Lovelock black holes in diverse dimensions,''
  Eur. Phys. J. C \textbf{74},  3074 (2014)
  [arXiv:1405.4143 [gr-qc]].
  %%CITATION = doi:10.1140/epjc/s10052-014-3074-1;%%
  %40 citations counted in INSPIRE as of 10 Feb 2017
%\cite{Sadeghi:2016dvc}
\bibitem{Sadeghi:2016dvc}
  J.~Sadeghi, B.~Pourhassan and M.~Rostami,
  %``P-V criticality of logarithm-corrected dyonic charged AdS black holes,''
  Phys. Rev. D \textbf{94}, 064006 (2016)
  [arXiv:1605.03458 [gr-qc]].
  %%CITATION = doi:10.1103/PhysRevD.94.064006;%%
  %6 citations counted in INSPIRE as of 10 Feb 2017
%\cite{Hansen:2016ayo}
\bibitem{Hansen:2016ayo}
  D.~Hansen, D.~Kubiznak and R.~B.~Mann,
  %``Universality of P-V Criticality in Horizon Thermodynamics,''
  JHEP \textbf{1701}, 047 (2017)
  [arXiv:1603.05689 [gr-qc]].
  %%CITATION = doi:10.1007/JHEP01(2017)047;%%
  %5 citations counted in INSPIRE as of 10 Feb 2017
%\cite{Lan:2017yia}
\bibitem{Lan:2017yia}
  S.~Lan and W.~Liu,
  %``Adiabatic Processes for Charged AdS Black Hole in the Extended Phase Space,''
  arXiv:1701.04662 [hep-th].  %%CITATION = ARXIV:1701.04662;%%
%\cite{Liang:2017rng}
\bibitem{Liang:2017rng}
  J.~Liang, Z.~H.~Guan, Y.~C.~Liu and B.~Liu,
  %``P–v criticality in the extended phase space of a noncommutative geometry inspired
  %Reissner–Nordström black hole in AdS space-time,''
  Gen. Rel. Grav. \textbf{49}, 29 (2017).
  %%CITATION = doi:10.1007/s10714-017-2189-8;%%
%\cite{Zeng:2016fsb}
\bibitem{Zeng:2016fsb}
  X.~X.~Zeng and L.~F.~Li,
  %``Holographic Phase Transition Probed by Nonlocal Observables,''
  Adv. High Energy Phys.  \textbf{2016}, 6153435 (2016)  [arXiv:1609.06535 [hep-th]].
  %%CITATION = doi:10.1155/2016/6153435;%%
%\cite{Hendi:2016usw}
\bibitem{Hendi:2016usw}
  S.~H.~Hendi, B.~Eslam Panah, S.~Panahiyan and M.~S.~Talezadeh,
  %``Geometrical thermodynamics and P-V criticality of black holes with power-law Maxwell field,''
  Eur. Phys. J. C \textbf{77},  133 (2017)
  [arXiv:1612.00721 [hep-th]].
  %%CITATION = doi:10.1140/epjc/s10052-017-4693-0;%%
  %3 citations counted in INSPIRE as of 27 Mar 2017
%\cite{Xu:2015rfa}
\bibitem{Xu:2015rfa}
  J.~Xu, L.~M.~Cao and Y.~P.~Hu,
 % ``P-V criticality in the extended phase space of black holes in massive gravity,''
  Phys. Rev. D \textbf{91}, 124033 (2015)
  [arXiv:1506.03578 [gr-qc]].
  %%CITATION = doi:10.1103/PhysRevD.91.124033;%%
  %40 citations counted in INSPIRE as of 16 Dec 2016
%\cite{Zou:2016sab}
\bibitem{Zou:2016sab}
  D.~C.~Zou, R.~Yue and M.~Zhang,
  %``Reentrant phase transitions of higher-dimensional AdS black holes in dRGT massive gravity,''
  Eur.\ Phys.\ J.\ C \textbf{77}, 256 (2017)
  arXiv:1612.08056 [gr-qc].
  %%CITATION = ARXIV:1612.08056;%%
  %1 citations counted in INSPIRE as of 10 Feb 2017
%\cite{Hendi:2017fxp}
\bibitem{Hendi:2017fxp}
  S.~H.~Hendi, R.~B.~Mann, S.~Panahiyan and B.~Eslam Panah,
  %``Van der Waals like behavior of topological AdS black holes in massive gravity,''
  Phys. Rev. D \textbf{95}, 021501 (2017)
  [arXiv:1702.00432 [gr-qc]].
  %%CITATION = doi:10.1103/PhysRevD.95.021501;%%
  %1 citations counted in INSPIRE as of 09 Feb 2017
%\cite{Ning:2016usb}
\bibitem{Ning:2016usb}
  S.~L.~Ning and W.~B.~Liu,
  %``Black Hole Phase Transition in Massive Gravity,''
  Int. J. Theor. Phys.  \textbf{55}, 3251 (2016).
  %%CITATION = doi:10.1007/s10773-016-2955-5;%%
%\cite{Prasia:2016fcc}
\bibitem{Prasia:2016fcc}
  P.~Prasia and V.~C.~Kuriakose,
  %``Quasi Normal Modes and P-V Criticallity for scalar perturbations
  %in a class of dRGT massive gravity around Black Holes,''
  Gen. Rel. Grav.  \textbf{48},  89 (2016)
  [arXiv:1606.01132 [gr-qc]].
  %%CITATION = doi:10.1007/s10714-016-2083-9;%%
  %3 citations counted in INSPIRE as of 09 Feb 2017
%\cite{Zeng:2015tfj}
\bibitem{Zeng:2015tfj}
  X.~X.~Zeng, H.~Zhang and L.~F.~Li,
  %``Phase transition of holographic entanglement entropy in massive gravity,''
  Phys. Lett. B \textbf{756}, 170 (2016)
  [arXiv:1511.00383 [gr-qc]].
  %%CITATION = doi:10.1016/j.physletb.2016.03.013;%%
  %18 citations counted in INSPIRE as of 23 Mar 2017

%\cite{Nollert:1999ji}
\bibitem{Nollert:1999ji}
  H.~P.~Nollert,
  %``TOPICAL REVIEW: Quasinormal modes: the characteristic `sound' of black holes and neutron stars,''
  Class. Quant. Grav. \textbf{16} (1999) R159.
  %%CITATION = doi:10.1088/0264-9381/16/12/201;%%
  %415 citations counted in INSPIRE as of 09 Feb 2017
%\cite{Kokkotas:1999bd}
\bibitem{Kokkotas:1999bd}
  K.~D.~Kokkotas and B.~G.~Schmidt,
  %``Quasinormal modes of stars and black holes,''
  Living Rev. Rel.  \textbf{2}, 2 (1999)
  [gr-qc/9909058].
  %%CITATION = doi:10.12942/lrr-1999-2;%%
  %654 citations counted in INSPIRE as of 09 Feb 2017
 %\cite{Konoplya:2011qq}
\bibitem{Konoplya:2011qq}
  R.~A.~Konoplya and A.~Zhidenko,
  %``Quasinormal modes of black holes: From astrophysics to string theory,''
  Rev. Mod. Phys.  \textbf{83}, 793 (2011)
  [arXiv:1102.4014 [gr-qc]].
  %%CITATION = doi:10.1103/RevModPhys.83.793;%%
  %247 citations counted in INSPIRE as of 09 Feb 2017
%\cite{Cardoso:2013pza}
\bibitem{Cardoso:2013pza}
  V.~Cardoso, ¨®.~J.~C.~Dias, G.~S.~Hartnett, L.~Lehner and J.~E.~Santos,
  %``Holographic thermalization, quasinormal modes and superradiance in Kerr-AdS,''
  JHEP \textbf{1404}, 183 (2014)   [arXiv:1312.5323 [hep-th]].
  %%CITATION = doi:10.1007/JHEP04(2014)183;%%  %48 citations counted in INSPIRE as of 23 Feb 2017
%\cite{Cardoso:2003cj}
\bibitem{Cardoso:2003cj}
  V.~Cardoso, R.~Konoplya and J.~P.~S.~Lemos,
  %``Quasinormal frequencies of Schwarzschild black holes in anti-de Sitter space-times:
  %A Complete study on the asymptotic behavior,''
  Phys. Rev. D \textbf{68}, 044024 (2003)  [gr-qc/0305037].
  %%CITATION = doi:10.1103/PhysRevD.68.044024;%%  %183 citations counted in INSPIRE as of 23 Feb 2017
%\cite{Warnick:2013hba}
\bibitem{Warnick:2013hba}
  C.~M.~Warnick,
  %``On quasinormal modes of asymptotically anti-de Sitter black holes,''
  Commun. Math. Phys.  \textbf{33}, 959 (2015) [arXiv:1306.5760 [gr-qc]].
  %%CITATION = doi:10.1007/s00220-014-2171-1;%%  %25 citations counted in INSPIRE as of 23 Feb 2017
%\cite{Konoplya:2002ky}
\bibitem{Konoplya:2002ky}
  R.~A.~Konoplya,
  %``Decay of charged scalar field around a black hole: Quasinormal modes of R-N, R-N-AdS black hole,''
  Phys. Rev. D \textbf{66}, 084007 (2002) [gr-qc/0207028].
  %%CITATION = doi:10.1103/PhysRevD.66.084007;%%  %115 citations counted in INSPIRE as of 23 Feb 2017
%\cite{Konoplya:2008rq}
\bibitem{Konoplya:2008rq}
  R.~A.~Konoplya and A.~Zhidenko,
  %``Stability of higher dimensional Reissner-Nordstrom-anti-de Sitter black holes,''
  Phys. Rev. D \textbf{78}, 104017 (2008)  [arXiv:0809.2048 [hep-th]].
  %%CITATION = doi:10.1103/PhysRevD.78.104017;%%  %59 citations counted in INSPIRE as of 23 Feb 2017
%\cite{Li:2016kws}
\bibitem{Li:2016kws}
  R.~Li, H.~Zhang and J.~Zhao,
  %``Time evolutions of scalar field perturbations in D-dimensional Reissner¨CNordstr?m Anti-de Sitter black holes,''
  Phys. Lett. B \textbf{758}, 359 (2016)  [arXiv:1604.01267 [gr-qc]].
  %%CITATION = doi:10.1016/j.physletb.2016.05.031;%%  %6 citations counted in INSPIRE as of 23 Feb 2017
%\cite{Rao:2007zzb}
\bibitem{Rao:2007zzb}
  X.~P.~Rao, B.~Wang and G.~H.~Yang,
  %``Quasinormal modes and phase transition of black holes,''
  Phys. Lett. B \textbf{649}, 472 (2007)
  [arXiv:0712.0645 [gr-qc]].
  %%CITATION = doi:10.1016/j.physletb.2007.04.049;%%
  %25 citations counted in INSPIRE as of 09 Feb 2017
%\cite{He:2010zb}
\bibitem{He:2010zb}
  X.~He, B.~Wang, R.~G.~Cai and C.~Y.~Lin,
  %``Signature of the black hole phase transition in quasinormal modes,''
  Phys. Lett. B \textbf{688}, 230 (2010)
  [arXiv:1002.2679 [hep-th]].
  %%CITATION = doi:10.1016/j.physletb.2010.04.006;%%
  %25 citations counted in INSPIRE as of 09 Feb 2017
%\cite{Berti:2008xu}
\bibitem{Berti:2008xu}
  E.~Berti and V.~Cardoso,
  %``Quasinormal modes and thermodynamic phase transitions,''
  Phys. Rev. D \textbf{77}, 087501 (2008)
  [arXiv:0802.1889 [hep-th]].
  %%CITATION = doi:10.1103/PhysRevD.77.087501;%%
  %15 citations counted in INSPIRE as of 09 Feb 2017
 %\cite{He:2008im}
\bibitem{He:2008im}
  X.~He, B.~Wang, S.~Chen, R.~G.~Cai and C.~Y.~Lin,
  %``Quasinormal modes in the background of charged Kaluza-Klein black hole with squashed horizons,''
  Phys. Lett. B \textbf{665}, 392 (2008)
  [arXiv:0802.2449 [hep-th]].
  %%CITATION = doi:10.1016/j.physletb.2008.06.038;%%
  %22 citations counted in INSPIRE as of 09 Feb 2017
%\cite{Miranda:2008vb}
\bibitem{Miranda:2008vb}
  A.~S.~Miranda, J.~Morgan and V.~T.~Zanchin,
  %``Quasinormal modes of plane-symmetric black holes according to the AdS/CFT correspondence,''
  JHEP \textbf{0811}, 030 (2008)
  [arXiv:0809.0297 [hep-th]].
  %%CITATION = doi:10.1088/1126-6708/2008/11/030;%%
  %36 citations counted in INSPIRE as of 09 Feb 2017
%\cite{Cai:2011qm}
\bibitem{Cai:2011qm}
  R.~G.~Cai, X.~He, H.~F.~Li and H.~Q.~Zhang,
  %``Phase transitions in AdS soliton spacetime through marginally stable modes,''
  Phys. Rev. D \textbf{84}, 046001 (2011)  [arXiv:1105.5000 [hep-th]].
  %%CITATION = doi:10.1103/PhysRevD.84.046001;%%  %15 citations counted in INSPIRE as of 25 Feb 2017
%\cite{Pan:2011hj}
\bibitem{Pan:2011hj}
  Q.~Y.~Pan and R.~K.~Su,
  %``Quasinormal Modes of Phantom Scalar Perturbation in Background of Reissner-Nordstrom Black Hole,''
  Commun. Theor. Phys.  \textbf{55}, 221 (2011).
  %%CITATION = doi:10.1088/0253-6102/55/2/07;%%
  %3 citations counted in INSPIRE as of 09 Feb 2017
%\cite{Shen:2007xk}
\bibitem{Shen:2007xk}
  J.~Shen, B.~Wang, C.~Y.~Lin, R.~G.~Cai and R.~K.~Su,
  %``The phase transition and the Quasi-Normal Modes of black Holes,''
  JHEP \textbf{0707}, 037 (2007)
  [hep-th/0703102 [HEP-TH]].
  %%CITATION = doi:10.1088/1126-6708/2007/07/037;%%
  %28 citations counted in INSPIRE as of 09 Feb 2017
  %\cite{Koutsoumbas:2006xj}
\bibitem{Koutsoumbas:2006xj}
  G.~Koutsoumbas, S.~Musiri, E.~Papantonopoulos and G.~Siopsis,
  %``Quasi-normal Modes of Electromagnetic Perturbations of Four-Dimensional Topological Black Holes with Scalar Hair,''
  JHEP \textbf{0610}, 006 (2006)
  [hep-th/0606096].
  %%CITATION = doi:10.1088/1126-6708/2006/10/006;%%
  %49 citations counted in INSPIRE as of 09 Feb 2017
%\cite{Zou:2014sja}
\bibitem{Zou:2014sja}
  D.~C.~Zou, Y.~Liu, C.~Y.~Zhang and B.~Wang,
  %``Dynamical probe of thermodynamical properties in three-dimensional hairy AdS black holes,''
  Europhys. Lett.  \textbf{116}, 40005 (2016)
  [arXiv:1411.6740 [hep-th]].
  %%CITATION = doi:10.1209/0295-5075/116/40005;%%
  %1 citations counted in INSPIRE as of 09 Feb 2017
%\cite{Zangeneh:2017rhc}
\bibitem{Zangeneh:2017rhc}
  M.~K.~Zangeneh, B.~Wang, A.~Sheykhi and Z.~Y.~Tang,
  %``Charged scalar quasi-normal modes for linearly charged dilaton-Lifshitz solutions,''
  arXiv:1701.03644 [hep-th].
  %%CITATION = ARXIV:1701.03644;%%
%\cite{Liu:2014gvf}
\bibitem{Liu:2014gvf}
  Y.~Liu, D.~C.~Zou and B.~Wang,
  %``Signature of the Van der Waals like small-large charged AdS black hole phase transition in quasinormal modes,''
  JHEP \textbf{1409}, 179 (2014)
  [arXiv:1405.2644 [hep-th]].
  %%CITATION = doi:10.1007/JHEP09(2014)179;%%
  %36 citations counted in INSPIRE as of 14 Jan 2017
%\cite{Chabab:2016cem}
\bibitem{Chabab:2016cem}
  M.~Chabab, H.~El Moumni, S.~Iraoui and K.~Masmar,
  %``Behavior of quasinormal modes and high dimension RN¨CAdS black hole phase transition,''
  Eur. Phys. J. C \textbf{76}, 676 (2016)
  [arXiv:1606.08524 [hep-th]].
  %%CITATION = doi:10.1140/epjc/s10052-016-4518-6;%%
  %5 citations counted in INSPIRE as of 09 Feb 2017
%\cite{Chabab:2017knz}
\bibitem{Chabab:2017knz}
  M.~Chabab, H.~El Moumni, S.~Iraoui and K.~Masmar,
  %``Phase Transition of Charged-AdS Black Holes and Quasinormal Modes : a Time Domain Analysis,''
  arXiv:1701.00872 [hep-th].
  %%CITATION = ARXIV:1701.00872;%%
%\cite{Mahapatra:2016dae}
\bibitem{Mahapatra:2016dae}
  S.~Mahapatra,
  %``Thermodynamics, Phase Transition and Quasinormal modes with Weyl corrections,''
  JHEP \textbf{1604}, 142 (2016)
  [arXiv:1602.03007 [hep-th]].
  %%CITATION = doi:10.1007/JHEP04(2016)142;%%
  %8 citations counted in INSPIRE as of 21 Mar 2017
%\cite{Alsup:2008fr}
\bibitem{Alsup:2008fr}
  J.~Alsup and G.~Siopsis,
  %``Low-lying quasinormal modes of topological AdS black holes and hydrodynamics,''
  Phys. Rev. D \textbf{78}, 086001 (2008)
  [arXiv:0805.0287 [hep-th]].
  %%CITATION = doi:10.1103/PhysRevD.78.086001;%%
  %11 citations counted in INSPIRE as of 01 May 2017
%\cite{Gonzalez:2012xc}
\bibitem{Gonzalez:2012xc}
  P.~A.~Gonzalez, F.~Moncada and Y.~Vasquez,
  %``Quasinormal Modes, Stability Analysis and Absorption Cross Section for
  %4-dimensional Topological Lifshitz Black Hole,''
  Eur. Phys. J. C \textbf{72}, 2255 (2012)
  [arXiv:1205.0582 [gr-qc]].
  %%CITATION = doi:10.1140/epjc/s10052-012-2255-z;%%
  %16 citations counted in INSPIRE as of 01 May 2017
%\cite{Becar:2012bj}
\bibitem{Becar:2012bj}
  R.~Becar, P.~A.~Gonzalez and Y.~Vasquez,
  %``Quasinormal modes and stability analysis for z=4 Topological black
  %hole in 4+1 dimensional Horava-Lifshitz gravity,''
  Int. J. Mod. Phys. D \textbf{22}, 1350007 (2013)
  [arXiv:1210.7561 [gr-qc]].
  %%CITATION = doi:10.1142/S0218271813500077;%%
  %10 citations counted in INSPIRE as of 01 May 2017
%\cite{Balazs:1986uj}
\bibitem{Balazs:1986uj}
  N.~L.~Balazs and A.~Voros,
  %``Chaos on the pseudosphere,''
  Phys. Rept.  \textbf{143}, 109 (1986).
  %%CITATION = doi:10.1016/0370-1573(86)90159-6;%%
  %161 citations counted in INSPIRE as of 01 May 2017
%\cite{Becar:2015kpa}
\bibitem{Becar:2015kpa}
  R.~Becar, P.~A.~Gonzalez and Y.~Vasquez,
  %``Quasinormal modes of Four Dimensional Topological Nonlinear Charged Lifshitz Black Holes,''
  Eur. Phys. J. C \textbf{76}, 78 (2016)
  [arXiv:1510.06012 [gr-qc]].
  %%CITATION = doi:10.1140/epjc/s10052-016-3937-8;%%
  %4 citations counted in INSPIRE as of 01 May 2017

\end{thebibliography}
\end{document}